\documentclass[sigplan,twocolumn]{acmart}
\renewcommand\footnotetextcopyrightpermission[1]{}
\usepackage{tikz}
\usepackage{amsmath}
\usepackage{graphicx}
\usepackage{filecontents}
\usepackage{booktabs} 
\usepackage{multicol} 
\usepackage{pifont}
\usepackage{algorithm}
\usepackage{algorithmic}
\usepackage{subcaption} 
\usepackage{caption} 
\AtBeginDocument{%
  }

\setcopyright{acmlicensed}
\copyrightyear{2026}
\acmYear{2026}
\acmConference[Preprint]{}
\acmISBN{under review}

\begin{document}

\title{Breaking the Boundaries of Long-Context LLM Inference: Adaptive KV Management on a Single Commodity GPU
}


 \author{He Sun}
 \affiliation{%
   \institution{ Department of Computer Science and Technology \& Suzhou Institute for Advanced Research,\\ University of Science and Technology of China }
   \city{Hefei}
   \country{China}}

 \author{Li Li}
 \affiliation{%
   \institution{IOTSC, University of Macau}
   \city{Macau}
   \country{China}
 }

 \author{Mingjun Xiao}
 \affiliation{%
  \institution{Department of Computer Science and Technology \& Suzhou Institute for Advanced Research,\\ University of Science and Technology of China }
  \city{Hefei}
  \country{China}}

 \author{Chengzhong Xu}
 \affiliation{%
   \institution{IOTSC, University of Macau}
   \city{Macau}
   \country{China}}




\renewcommand{\shortauthors}{He Sun et al.}

\begin{abstract}
Advanced Large Language Models (LLMs) have achieved impressive performance across a wide range of complex and long-context natural language tasks. However, performing long-context LLM inference locally on a commodity GPU (a PC) with privacy concerns remains challenging due to the increasing memory demands of the key-value (KV) cache. Existing systems typically identify important tokens and selectively offload their KV data to GPU and CPU memory. The KV data needs to be offloaded to disk due to the limited memory on a commodity GPU, but the process is bottlenecked by token importance evaluation overhead and the disk’s low bandwidth. In this paper, we present LeoAM, the first efficient importance-aware long-context LLM inference system for a single commodity GPU with adaptive hierarchical GPU-CPU-Disk KV management. Our system employs an adaptive KV management strategy that partitions KV data into variable-sized chunks based on the skewed distribution of attention weights across different layers to reduce computational and additional transmission overheads. Moreover, we propose a lightweight KV abstract method, which minimizes transmission latency by storing and extracting the KV abstract of each chunk on disk instead of the full KV data. LeoAM also leverages the dynamic compression and pipeline techniques to further accelerate inference. Experimental results demonstrate that LeoAM achieves an average inference latency speedup of $3.46\times$, while maintaining comparable LLM response quality. In scenarios with larger batch sizes, it achieves up to a $5.47\times$ speedup.
\end{abstract}


%
\maketitle
\section{Introduction}
Large Language Models (LLMs) have demonstrated remarkable capabilities of handling complex natural language processing tasks \cite{zhao2023survey,cellstrat2023,keary2024,daivi2024} and power advanced applications such as multi-round dialogues and long document summarization\cite{wang2024leave,ma2024mmlongbench}, etc. These applications often require processing large amounts of contextual tokens. Fortunately, the LLM models developed by industries have shown notable performance for processing long contexts as shown in Figure \ref{fig:ContextSizeModel} (a). Meanwhile, due to privacy concerns and reduced inference costs, there is an emerging trend to directly run those tasks on personal computing platforms, e.g., a PC equipped with a single commodity GPU \cite{privategpt2024,tuggener2024so,zhang2024instruction,zhang2024enabling,li2024personal,song2023powerinfer}. 
 
Long-context LLM inference on a single local GPU remains challenging. In addition to storing model weights, the memory usage of the KV cache increases linearly with the context length \cite{pope2023efficiently,beltagy2020longformer}, making limited memory a critical bottleneck. Fortunately, recent research \cite{zhang2023h2o,tangquest,zhao2024alisa,lee2024infinigen,zhang2024pqcache,liu2024scissorhands} indicates substantial sparsity in long context inference, i.e., identifying important tokens' KV and evicting the less important KV data, which can also preserve the output quality of the LLM. They cache the KV data of important tokens in the GPU for attention computation, and optimize hybrid GPU-CPU KV data offloading to reduce latency. 

However, performing long-context LLM inference on a resource-constrained commodity GPU poses many new challenges. First, the memory capacity of a commodity GPU is insufficient for LLM inference as the context length increases. For example, inferring a 64K context size using LLaMA-7B model requires approximately 47GB of memory containing 33.5G KV Cache, which far exceeds the memory limits of a commodity GPU and CPU (shown as Figure \ref{fig:ContextSizeModel} (b)). This necessitates caching KV data to external disk, in order to accommodate lengthy sequences or large batches (see details in \$ \ref{Sec:Long-Context}). Unfortunately, numerous studies have shown that token importance is query-dependent \cite{lee2024infinigen,tangquest,chen2025impress}, requiring frequent I/O between queries in token-importance-aware LLM inference frameworks. The limited I/O bandwidth of the disk becomes a significant performance bottleneck.

To cope with the above scenario and leverage the inherent sparsity of long-context inputs, we propose LeoAM, a token-importance-aware LLM inference acceleration system for a single commodity GPU, which utilizes GPU memory, CPU memory, and disk for hierarchical offloading of KV data to reduce the inference latency. The core idea of LeoAM is to efficiently identify and offload only the important tokens’ KV data by employing lightweight KV management, thereby reducing the immense disk loading latency. Nevertheless, building such an efficient LLM inference system with GPU-CPU-Disk KV data offloading involves two key challenges.

First, existing token importance evaluation methods often incur significant computational and transmission overhead due to the static KV management. On the one hand, many existing approaches perform importance evaluation at the token level \cite{zhang2023h2o,lee2024infinigen,zhao2024alisa,liu2024scissorhands}. When the context length grows, the number of tokens involved in the evaluation increases, resulting in considerable computational overhead. On the other hand, some approaches merge consecutive tokens into chunks \cite{gao2024cost,tangquest,juravsky2024hydragen,ye2024chunkattention,zheng2024sglang} and evaluate their importance at the chunk level to reduce evaluation cost. However, not all tokens within a chunk share the same level of importance, e.g., unimportant tokens may be included in chunks identified as important, which leads to unnecessary transmission. 

Second, the low bandwidth of the disk becomes a major I/O bottleneck during token importance evaluation, significantly increasing the evaluation overhead. It will typically require loading all tokens’ KV data into GPU /CPU memory for evaluation if existing approaches \cite{zhang2023h2o,tangquest,zhao2024alisa,lee2024infinigen,liu2024scissorhands,zhang2024pqcache} are directly applied in our resource-constrained environment. This means that the KV data stored on the disk also needs to be transmitted and evaluated, which introduces substantial transmission latency under the GPU-CPU-Disk KV offloading framework. 

To tackle the first challenge, we analyze the attention weights (serving as a proxy for token importance) of the KV data from all previous tokens under each decoding query. We observe that the attention weight distribution is highly skewed: a large number of contiguous regions exhibit near-zero weights, while only a few small regions concentrate the high weights; moreover, the degree of skewness varies across different layers. Based on these insights, we adaptively partition the KV data into chunks of different sizes based on the sparsity to accurately extract the important tokens. Additionally, we dynamically adjust the chunk size across different layers according to their respective skewness levels. 

To address the second challenge, we aim to minimize the amount of KV data transmitted from the slower disk during token importance evaluation. To achieve this, we propose a lightweight KV abstract method. Building upon adaptive KV management, we extract a compact abstract from each chunk of KV data stored on disk. Specifically, we store only the parts of each chunk that are necessary for importance evaluation. During the importance evaluation, only these small-sized abstracts need to be loaded instead of the full KV data from disk. Meanwhile, we dynamically compress the KV data to transmit and pipeline the computation and transmission for further reducing inference latency.   

We implement LeoAM based on an offloading-based LLM inference system \cite{sheng2023flexgen} and evaluate it on a PC with different GPU memory specifications, using various models of different sizes and multiple datasets. Experimental results demonstrate that LeoAM achieves an average inference latency speedup of $3.46\times$, while maintaining comparable LLM response quality. In scenarios with larger batch sizes, it achieves up to a $5.47\times$ speedup.

The main contributions are summarized as follows:

(1) We present LeoAM, the first efficient importance-aware long-context LLM inference system based on the GPU-CPU-Disk offloading designed for a single commodity GPU.

(2) We propose an adaptive KV management to reduce both the computational and transmission overheads during importance evaluation.

(3) We design a KV abstract and compression strategy to reduce latency caused by the disk’s limited bandwidth. Moreover, we also proposed a dynamic three-tier pipeline method to further reduce the inference latency.

(4) Finally, we implement LeoAM based on an offloading-based LLM inference system, and the evaluation shows that our system can achieve an average inference latency speedup of $3.46\times$ compared to the state-of-the-art system while maintaining the LLM output quality.

\begin{figure}[htbp]
    \centering
    \includegraphics[width=1.02\linewidth]{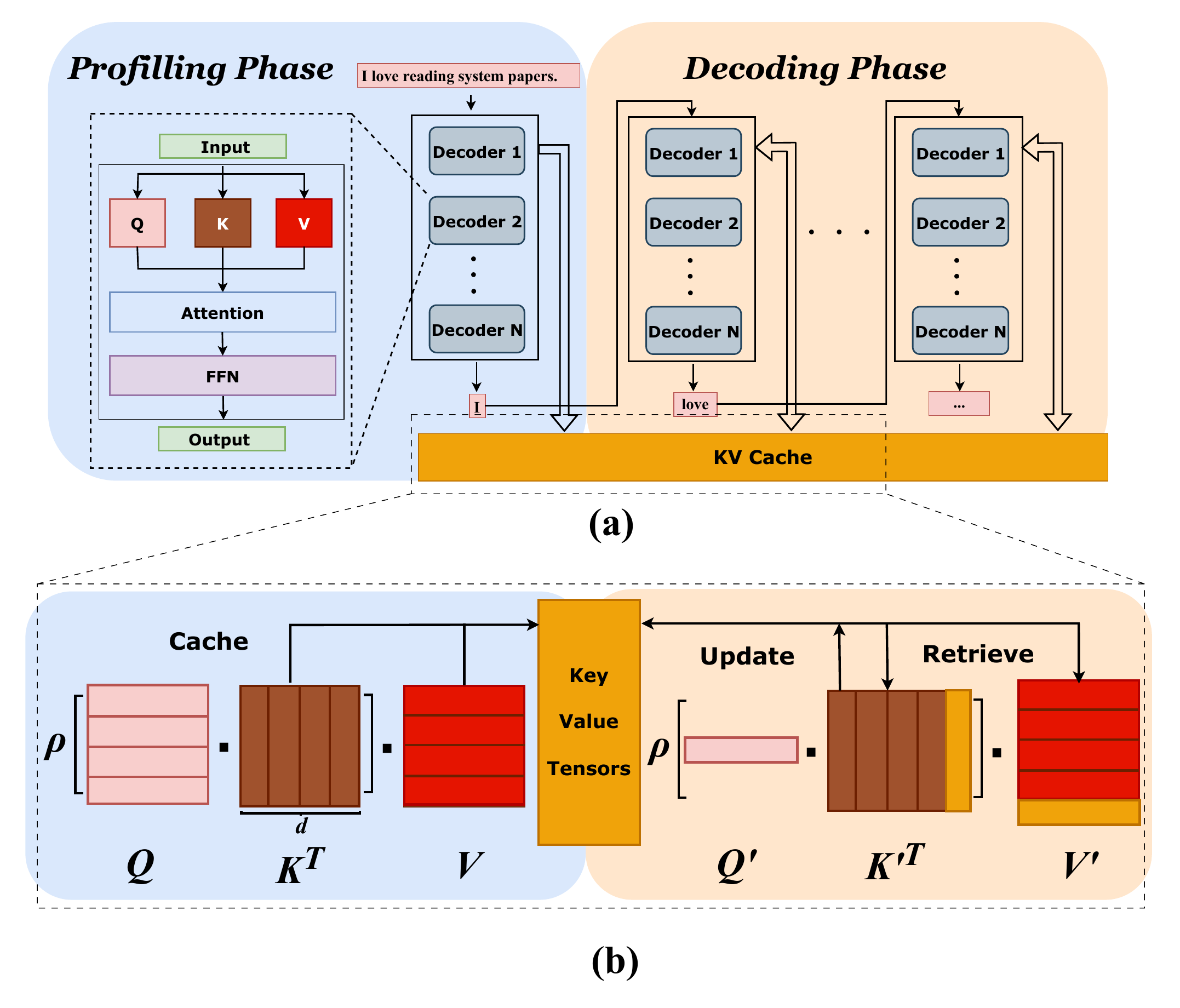}
    \caption{The LLM Inference Workflow and KV Caching.}
    \label{fig:LLL_iner_load}
\end{figure}

\section{Background}\label{Sec: B_M}
\subsection{Large Language Model Inference Basics}
The majority of Large Language Models (LLMs) are built on the decoder-only transformer architecture and are composed of a large stack of decoder blocks \cite{vaswani2017attention,chowdhery2023palm,brown2020language}. 
\begin{figure}[tbp]
    \begin{minipage}[t]{0.50\linewidth}
        \centering
        \includegraphics[width=\textwidth]{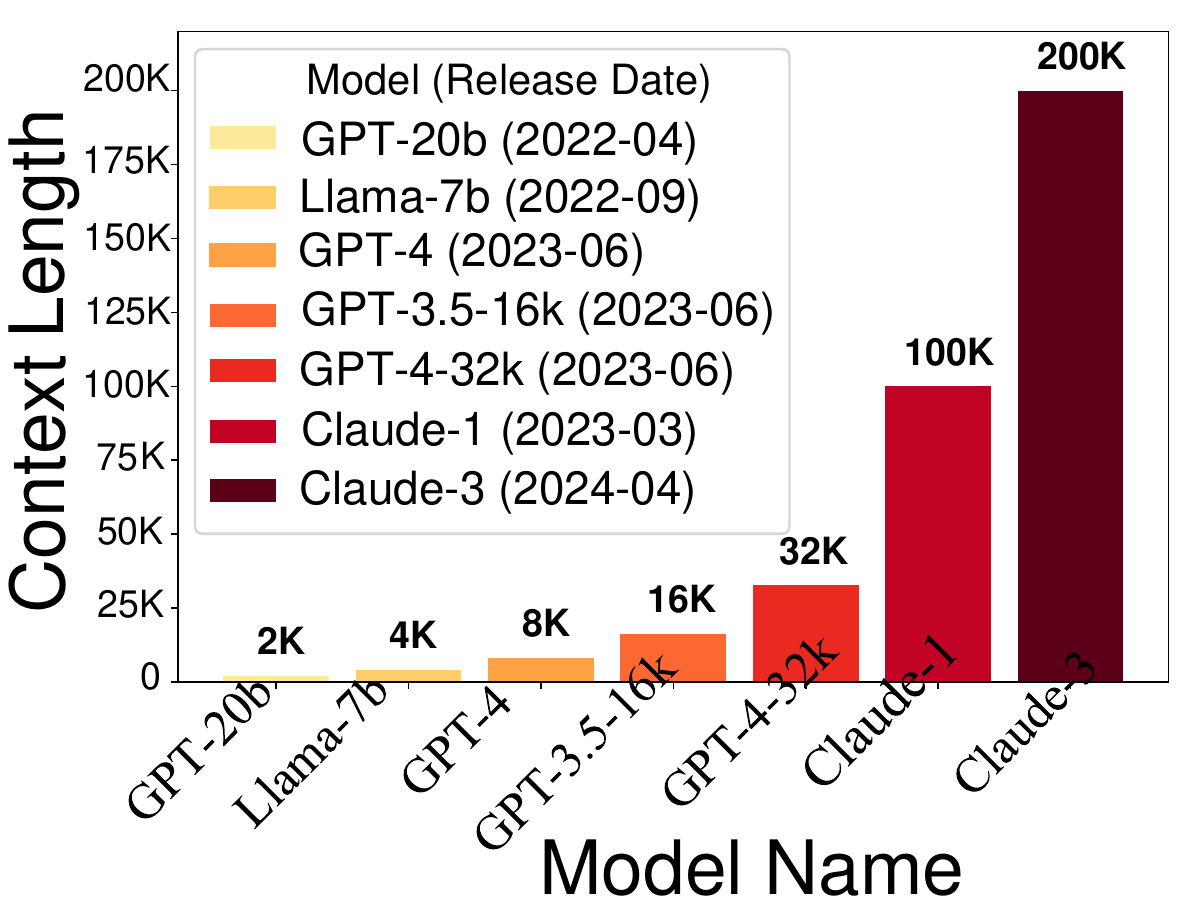}
        \centerline{(a) Models Context Length.}
    \end{minipage}%
    \begin{minipage}[t]{0.50\linewidth}
        \centering
        \includegraphics[width=\textwidth]{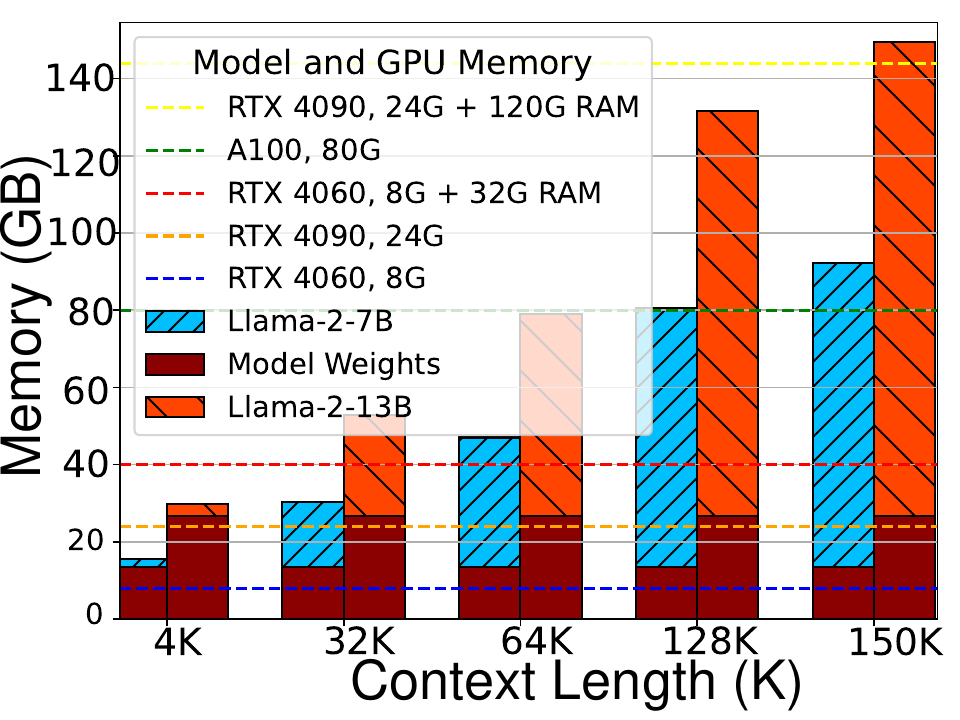}
        \centerline{(b) Context Length VS. Memory.}
    \end{minipage}
    \caption{Long-Context LLM Inference. The left figure shows the growth trend of the supported context length of LLM. The right figure shows the memory footprint of different models and context lengths during inference and the memory capacities of common devices.}
    \label{fig:ContextSizeModel}
\end{figure}

\noindent\textbf{LLM Inference Workflow.}
During inference, the process is typically divided into two stages: prefilling and decoding phrases \cite{zhou2024survey}, as shown in Figure \ref{fig:LLL_iner_load} (a). In the prefilling phase, the text input for LLM inference is first tokenized (a single word is considered one token typically). For instance, the input “I love reading system papers” shown in Figure \ref{fig:LLL_iner_load} (a) is converted into 5 tokens (excluding the start symbol <S> in some LLMs). The LLM then processes all prompt tokens in one pass. In the decoding phase, LLMs generate output tokens one by one in an autoregressive manner, where each newly generated token, along with all previous tokens, is fed back into the model to predict the next token. Each output token generation requires access to the entire set of LLM parameters during the decoding phase. The decoding phase continues until either the maximum output length is reached or an end-of-sequence '<EOS>' token is generated.

\noindent\textbf{Transformer-based Decoder.} During LLM inference, each transformer-based Decoder layer consists mainly of Query, Key, and Value vectors, self-attention (abbreviated as “attention”), and a feed-forward network (FFN), as shown on the left side of Figure \ref{fig:LLL_iner_load}(a). For a prompt input tokens sequence $X = \{x_1,x_2,...,x_n\}$, through GeMM computation, each transformer-based Decoder linearly transforms $X$ to obtain $Q$, $K$, and $V$ matrices using model parameter matrices. Subsequently, the semantic context of the prompt input is captured through attention computation and the attention score is computed by Equation (\ref{eq:attention}).
\begin{equation}\label{eq:attention}
    Attention(Q,K,V) = \rho(\frac{Q\cdot K^T}{\sqrt{d}})\cdot V
\end{equation}
where $\rho(\cdot)$ is the softmax function and $d$ is the hidden dimension. Further to enhance model capacity, Multi-Head Attention (MHA) enables multiple self-attention heads to process and capture contextual information from different perspectives in parallel\cite{vaswani2017attention}. Finally, the attention output is passed to the FFN to compute the results, which are then passed to the next Decoder.

\noindent\textbf{KV Caching Technology.} During the attention operation, the computational complexity is quadratic according to Equation (\ref{eq:attention}), which significantly increases computation latency. The overhead becomes more severe as context length increases and model size grows \cite{zhao2024alisa,beltagy2020longformer,tangquest}. A viable solution to this issue is KV Caching technology, which stores intermediate key ($K$) and value ($V$) pairs across different layers as KV Cache for reuse during the autoregressive generation of subsequent tokens. Figure \ref{fig:LLL_iner_load}(b) illustrates the KV Caching process in detail, which transforms the original quadratic complexity of general matrix multiplication (GeMM) into the general matrix-vector multiplication (GeMV) with linear complexity in memory access and significantly improves performance. 

\section{Motivations and Challenges}
 \begin{figure}[t]
    \begin{minipage}[t]{0.48\linewidth}
        \centering
        \includegraphics[width=\textwidth]{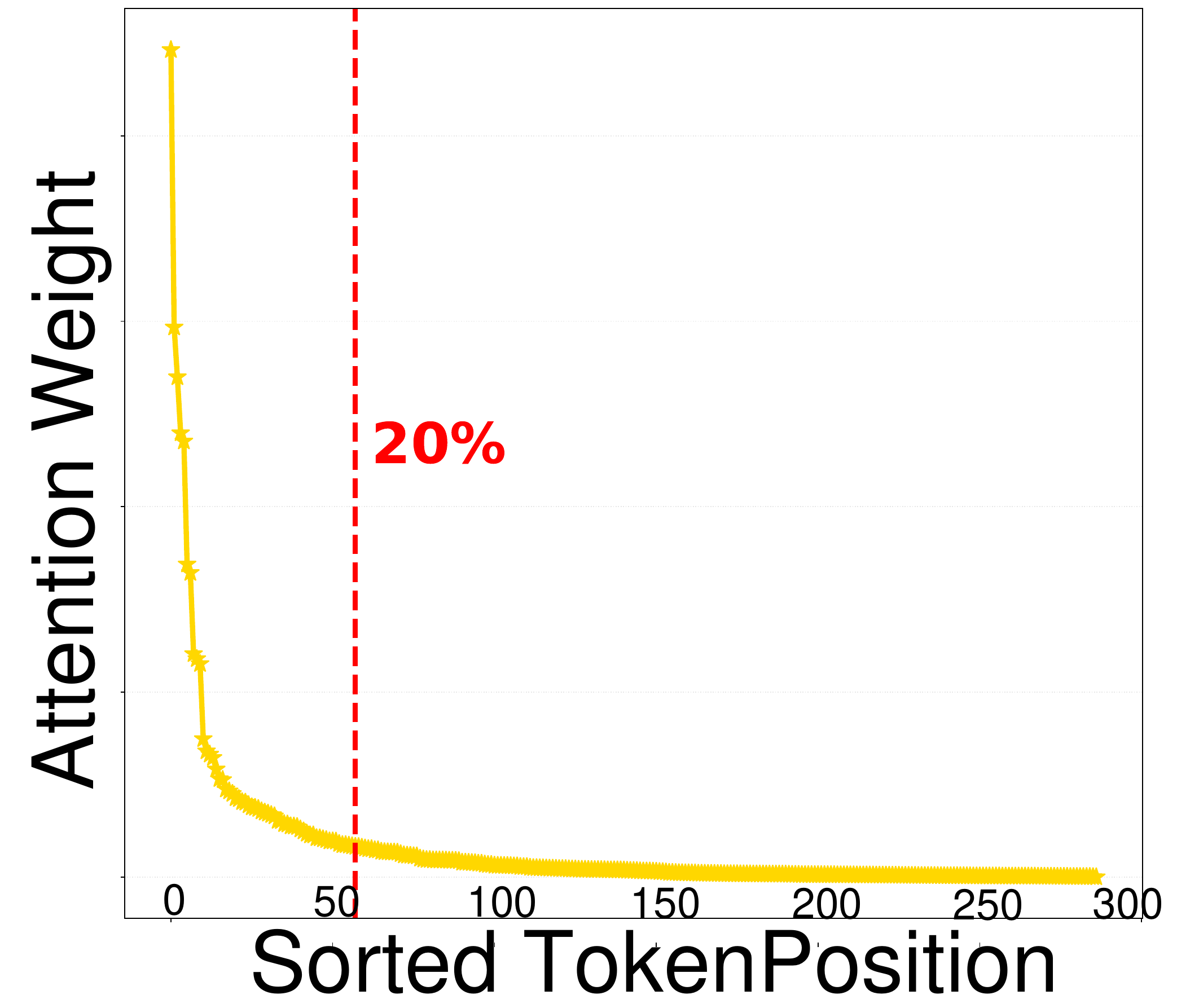}
        \centerline{(a) Tokens' sparsity.}
    \end{minipage}
    \begin{minipage}[t]{0.48\linewidth}
        \centering
        \includegraphics[width=\textwidth]{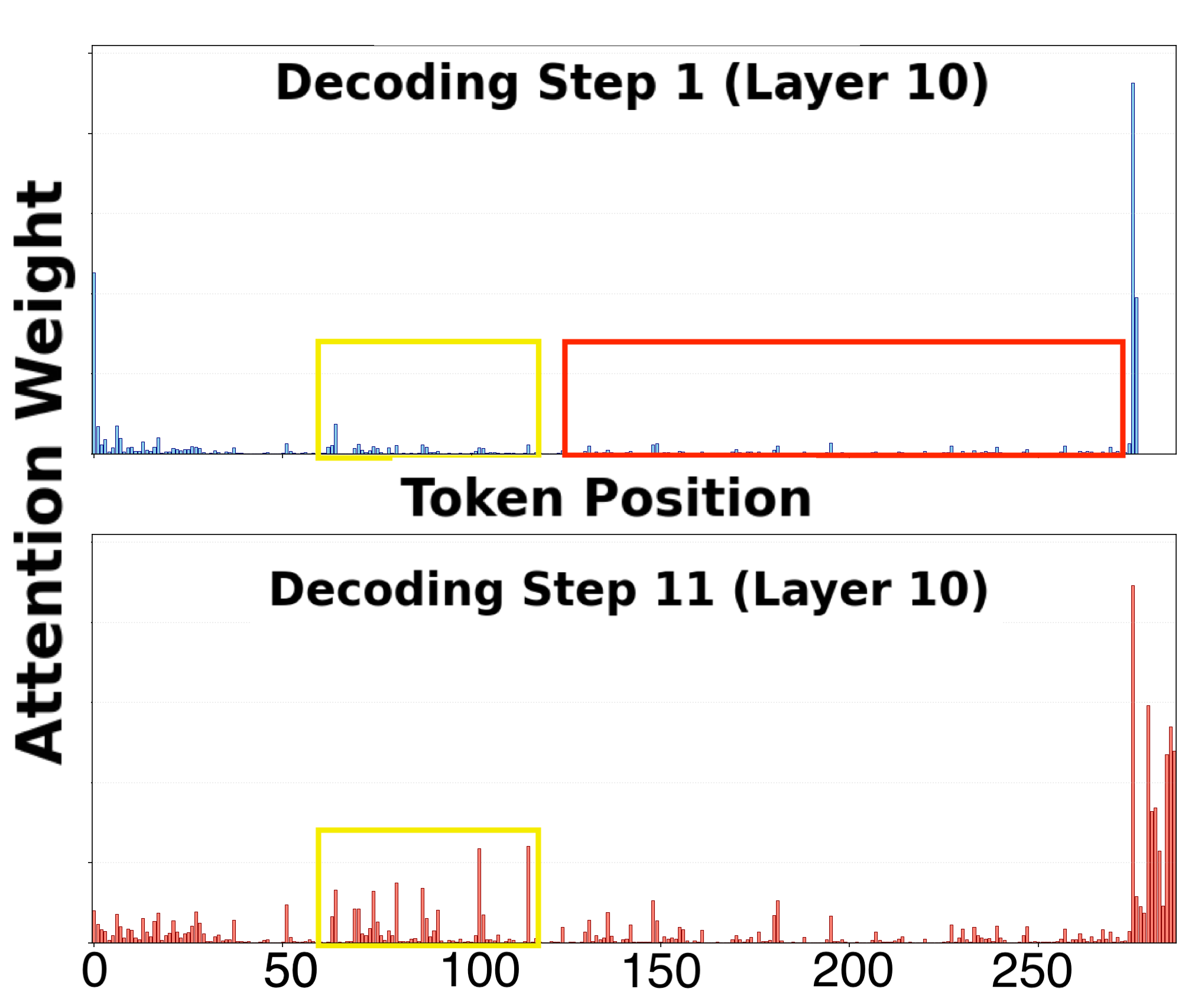}
        \centerline{(b) Tokens' attention weight.}
    \end{minipage}%
    \caption{The properties of Tokens. The left figure shows that only a small proportion (20\% in the figure) of KV tensors plays a dominant role in attention computation. The right figure shows the attention weight of different tokens in different decoding rounds.}
    \label{fig:token_analysis}
    \vspace{-0.1 in}
\end{figure}
\vspace{0.1in}
\subsection{The Long-Context LLM Inference}\label{Sec:Long-Context}
In recent years, long-context LLM inference has garnered significant attention. As shown in Figure \ref{fig:ContextSizeModel}(a), many companies have released models with increasingly large context windows, with some, like Claude-3\cite{anthropic2024claude3}, supporting up to 2 million context inputs. 
During LLM inference, the memory footprint of the KV Cache increases as the context length grows. As shown in Figure \ref{fig:ContextSizeModel}(b), when the context length\footnote{These are many fine-tuned Llama series models supporting larger context length. e.g., Yarn-Llama-2-13b\cite{YarnLlama2-13B} and 
Yarn-Llama-2-7B\cite{YarnLlama2-7B}} reaches 32K, the memory footprint for Llama-2-7B inference exceeds the GPU memory limit of an RTX 4090, a high-end commodity GPU. Similarly, at a context length of 64K, the memory demand surpasses the combined capacity of both GPU and CPU memory on a computer equipped with an RTX 4060, a low-end commodity GPU. When the context length reaches 150K, the memory required for Llama-2-13B inference exceeds the combined capacity of a high-end commodity RTX 4090 GPU and 120GB of CPU RAM. Furthermore, the batch size in Figure \ref{fig:ContextSizeModel} is set as 1. As the batch size increases, the memory usage of the KV Cache also grows linearly. In this scenario, we may need to cache additional KV data to external storage, such as an SSD (see more details in \$ \ref{Sec:Necessity}).

Summarily, as the context length grows, the KV Cache memory footprint scales linearly, rapidly consuming available memory resources. This imposes a critical bottleneck, particularly for LLM inference on a resource-constrained commodity GPU, necessitating efficient memory management strategies to enable scalable long-context LLM Inference.
\begin{figure}[tbp]
    \centering
    \includegraphics[width=0.45\textwidth]{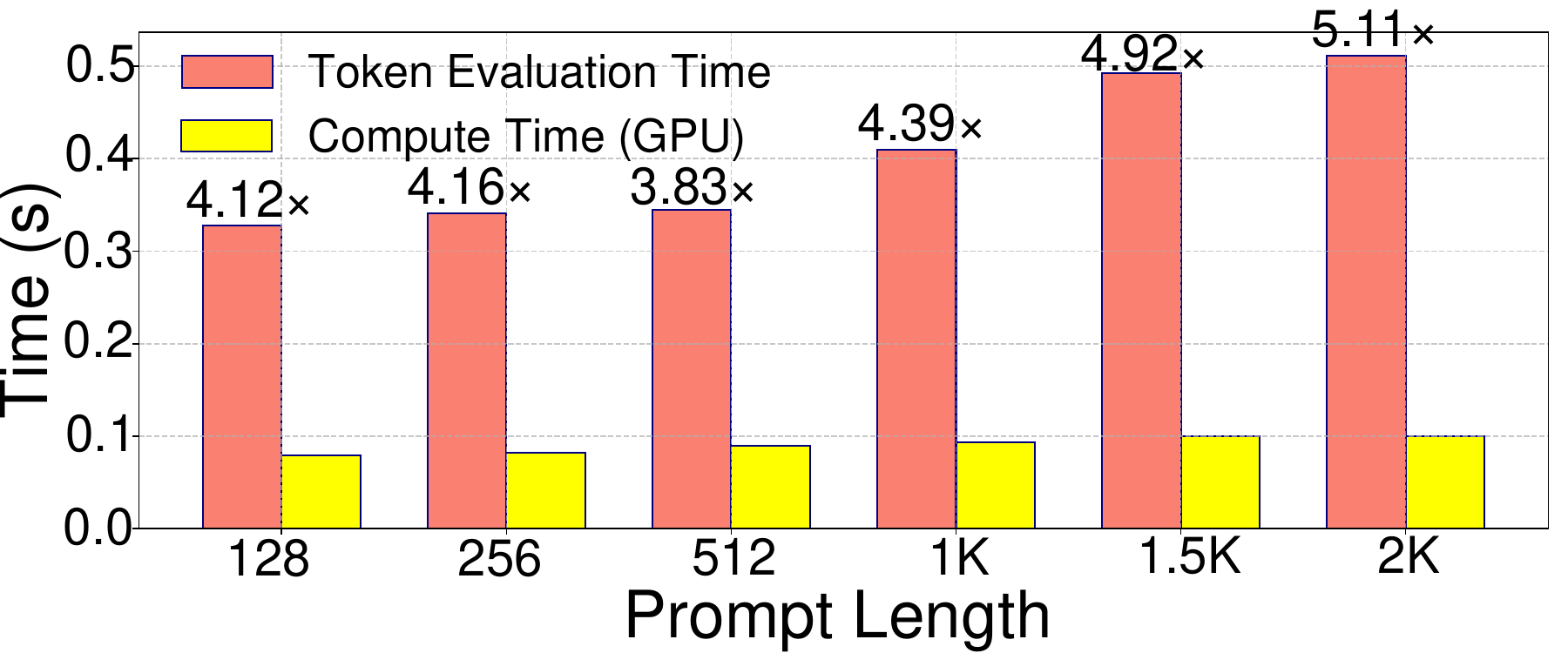}
    \caption{The Overhead of Token Evaluation in Token Level: Using OPT-2.7b to evaluate the token importance evaluation latency overhead and computation latency on LongBench.}
    \label{fig:Overhead}
\end{figure}
\begin{figure}[tbp]
    \centering
    \includegraphics[width=0.45\textwidth]{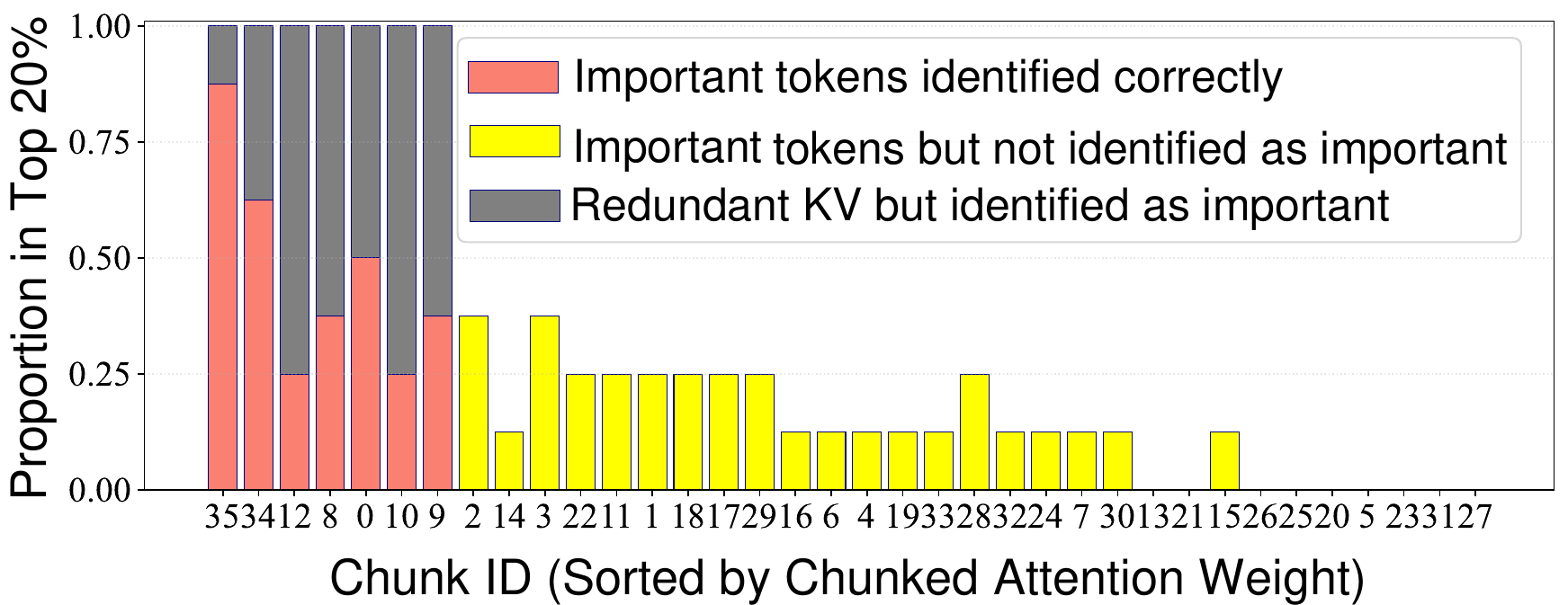}
    \caption{Proportion of Tokens in Top 20\% per Chunk/Page: The red part represents the important KV identified correctly; Grey part represents the redundant KV but identified as important.}
    \label{fig:proportions}
\end{figure}

\subsection{The Opportunity of Long-Context Sparsity}
Inspired by linguistic principles and transformer-based language models \cite{beltagy2020longformer, levy2014linguistic}, Long-context inputs exhibit a high degree of sparsity. As shown in Figure \ref{fig:token_analysis} (a), fewer than 20\% of the tokens exhibit high attention weights. Therefore, only a subset of KV tensors plays a dominant role in attention computation. The methods of recent works\cite{lee2024infinigen,tangquest,zhang2023h2o,liu2024scissorhands, zhao2024alisa} cache all KV data during the prefilling stage and then identify important tokens during the decoding stage at the token or chunk/page level. The KV data of important tokens is stored on the GPU, while less important tokens are offloaded to the CPU and discarded. This approach reduces computational overhead during the decoding stage while maintaining inference accuracy. 

Based on this opportunity, we can accelerate inference by offloading the KV data of tokens according to their importance. However, performing long-context inference under memory-constrained environments requires further system-level optimizations.

\begin{figure}[tbp]
    \centering
    \includegraphics[width=0.45\textwidth]{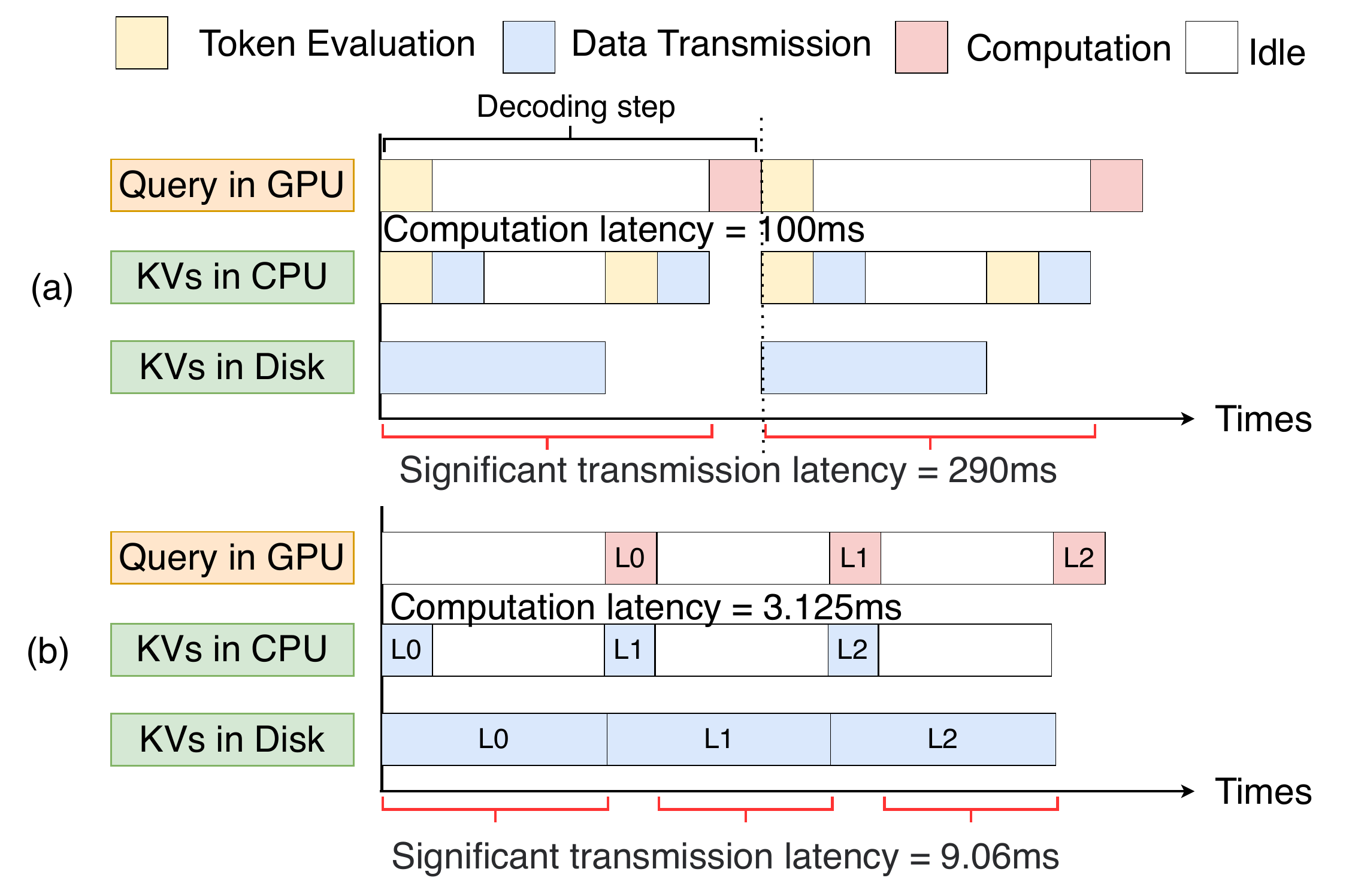}
    \caption{The I/O bottleneck of Data Transmission for Token Evaluation in GPU-CPU-Disk Offloading.}
    \label{fig:latency}
\end{figure}
\subsection{Memory-constrain Long-Context KV Offloading}\label{Sec:Necessity}
In recent works \cite{tangquest,zhao2024alisa,zhang2024pqcache,liu2024scissorhands,gemodel,chen2025impress,lee2024infinigen}, they have focused on GPU-CPU hierarchical KV Cache offloading on industrial-grade GPUs (e.g., A6000, V100). However, on a commodity GPU, KV Cache often needs to be offloaded to disk, such as an SSD. This necessity arises for two main reasons: \textit{first}, the memory-constrained commodity GPU cannot store such huge size of KV
 data (as shown in Figure \ref{fig:ContextSizeModel}(b) in \$ \ref{Sec:Long-Context}); \textit{second}, these GPU-CPU offloading approaches fail to address the issue that the KV data needs to be stored on disk. Meanwhile, H2O-like approaches \cite{zhang2023h2o} assume that the attention pattern remains consistent across iterations, which often breaks down in practice. Figure \ref{fig:token_analysis}(b) illustrates this phenomenon by showing the attention weights between previously generated tokens and the currently generated token at the $1^{st}$ and $11^{th}$ decoding steps. The yellow-highlighted region shows tokens whose KV Cache was unimportant in the $1^{st}$ decoding step and offloaded to disk, but became important in the $11^{th}$ step and had to be reloaded into the GPU for attention computation. 
 
Therefore, these existing KV cache offloading strategies cannot be directly applied to a single commodity GPU due to memory and I/O constraints in the long-context scenario. Building on this gap and the opportunity of long-context sparsity, we propose LeoAM, a long-context inference framework designed specifically for the commodity GPU. However, due to resource constraints, LeoAM still faces many challenges. 

\subsection{Challenges}\label{Sec:challenges}
\noindent \textbf{Challenge 1: Existing token importance evaluation methods for long-context inputs impose additional significant computation and transmission overhead on the commodity GPU.} On one hand, some works \cite{lee2024infinigen,zhang2023h2o,zhao2024alisa,liu2024scissorhands} perform importance evaluation/prediction at the token level. However, as the context length increases, the overhead of these evaluations/predictions becomes a bottleneck in resource-constrained environments. We test the token evaluation time and compute time of H2O\cite{zhang2023h2o} with various-length prompts. As shown in Figure \ref{fig:Overhead}, we find that the overhead of token importance evaluation is approximately 4 to 5 times the GPU computation time, representing a significant performance burden for H2O-like token-level methods. On the other hand, some other works perform token importance evaluation at the chunks/pages level \cite{gao2024cost,tangquest,juravsky2024hydragen,ye2024chunkattention,zheng2024sglang}. Nevertheless, these methods have two main problems. \textit{First}, chunk/page-based methods may mistakenly treat unimportant tokens as important, leading to unnecessary data transmission. As shown in Figure \ref{fig:proportions}, the black part of the KV data is originally redundant, but is evaluated in the top 20\% important chunk, which causes 40\% unnecessary data transmission. \textit{Second}, due to the important tokens constitute only a small proportion, some truly important tokens’ KV data may fail to be loaded into the GPU for attention computation, resulting in a lower recall rate and increased perplexity. Figure \ref{fig:proportions} shows that the yellow part of the KV data is important but hidden within chunks. 
\begin{figure}[tbp]
    \centering
    \includegraphics[width=0.45\textwidth]{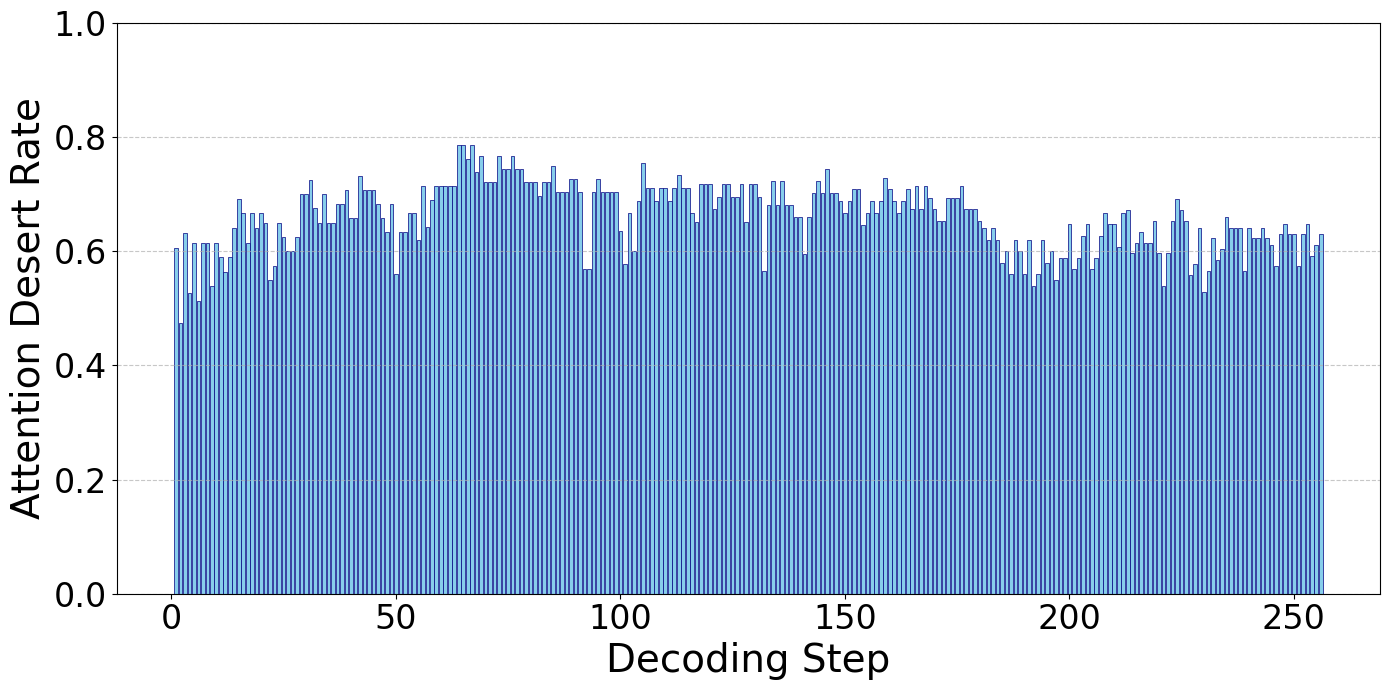}
    \caption{The attention desert rate across decoding steps.}
    \label{fig:attention_desert}
\end{figure}

\noindent \textbf{Challenge 2: Token-importance-aware KV data management with GPU-CPU-Disk offloading on the commodity GPU can lead to significant I/O bottlenecks.} During each decoding step, importance-aware KV data offloading is required; however, this process is often constrained by the low I/O bandwidth of the disk. On one hand, when evaluating token importance, tokens stored in the disk also need to be considered (\$\ref{Sec:Necessity}), and these KV data need to be first transmitted from disk to CPU memory for evaluation and then selected for loading into GPU memory for further LLM computation. During this process, significant transmission latency is incurred. We evaluate the latency when the context length is 2K, the batch size is 4, and 40\% of the KV cache is stored on disk. As shown in Figure \ref{fig:latency} (a), in a decoding step, although the computation latency is only 100 ms, the transmission latency reaches 290 ms, making disk transfer the primary bottleneck. On the other hand, when incorporating inter-layer prefetching techniques like \cite{lee2024infinigen}, the prefetching latency in each layer is 9.06 ms and often cannot be effectively hidden behind the computation latency of 3.125 ms in the GPU-CPU-Disk offloading due to the low bandwidth of the disk, as shown in figure \ref{fig:latency} (b).

\subsection{Insights of LeoAM}
\noindent\textbf{Insight 1:} In a given decoding step with the current query, the attention distribution is skewed and there often exists a contiguous region of preceding tokens with consistently low attention weights, which we call \textit{attention desert}. 

As shown in Figure \ref{fig:token_analysis} (b), the region highlighted in red represents a large attention desert. To demonstrate the distribution of \textit{attention desert}, we experimentally illustrate this insight using LongChat-7B-32K \cite{longchat7b} on the LongBench\cite{bai2023longbench} dataset. First, we define the "attention desert rate" as the proportion of unimportant chunks among all chunks. These chunks form what we call the "attention desert", and a higher value indicates a higher density of such regions. Figure \ref{fig:attention_desert} shows the attention desert rate across different decoding steps with a $10\%$ importance rate, $256$ tokens output, and a chunk size of $16$. The attention desert rate can achieve $60\%\sim80\%$, which indicates that a large proportion of consecutive regions consist of unimportant chunks. Due to the widely acknowledged token sparsity, the “attention desert” is consistently observed across a broad range of models and datasets. We leverage this observation to adjust chunk sizes adaptively: continuous unimportant chunks can be merged into a larger chunk and evaluated only once, while important chunks can be further split to identify and exclude unimportant sub-chunks. 

\begin{figure}[tbp]
    \centering
    \includegraphics[width=0.45\textwidth]{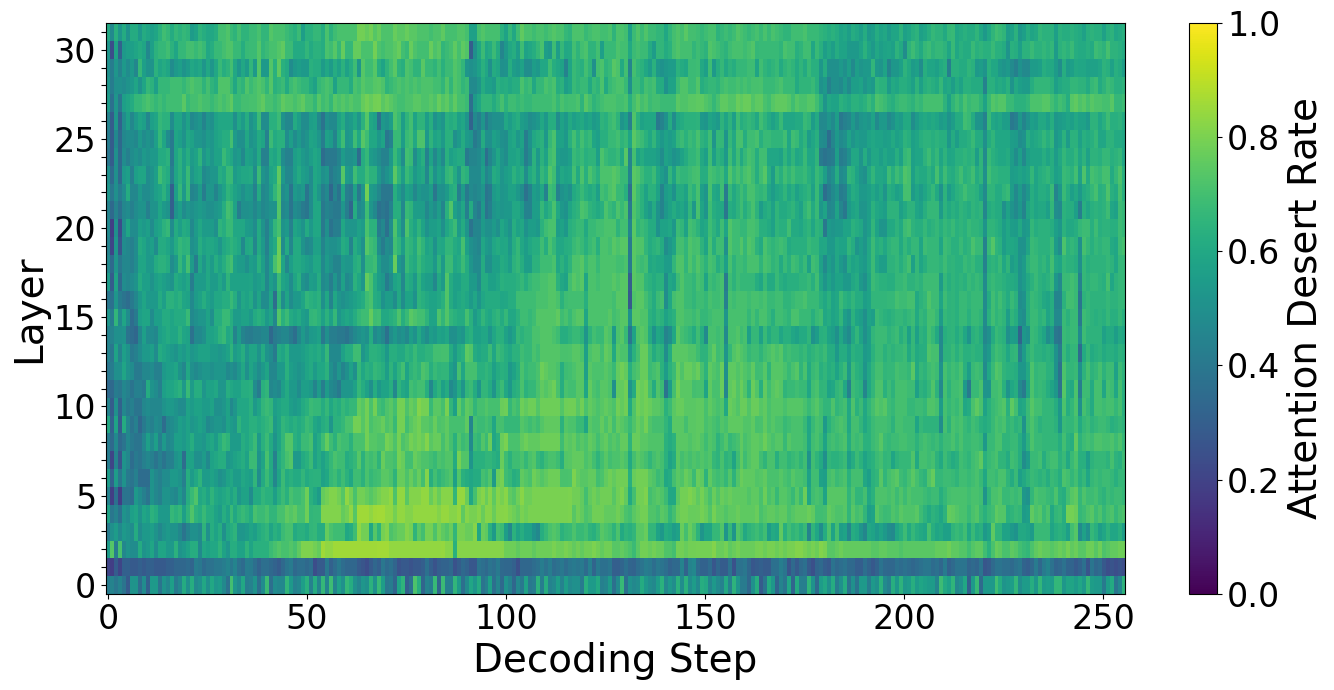}
    \caption{The attention desert rate heatmap across layers and decoding steps.}
    \label{fig:attention_desert_headmap}
\end{figure}
\noindent\textbf{Insight 2:} The attention desert rate varies across different layers of the LLM and different decoding layers.

To gain further observations, we conduct a detailed analysis of the attention desert rate across layers with the same experimental settings as Insight 1. Figure \ref{fig:attention_desert_headmap} shows the attention desert rate heatmap across layers and decoding steps. We observe that: (1) in the first two layers, the attention desert rate is significantly lower compared to adjacent layers; (2) during the initial decoding steps, approximately the first 5\% to 10\%, the attention desert rate also remains relatively low. Therefore, in these layers and decoding steps, we can resize the initial chunk size based on the distribution of the attention desert rate, enabling finer-grained management of the corresponding KV data.

\section{LeoAM Design}
In this section, we propose LeoAM, an efficient long-context LLM inference system for a commodity GPU. We first show the high-level overview of LeoAM and the basic workflow of such importance-informed offloading system (Section \ref{Sec:overview}) and discuss the importance-aware adaptive KV management (Section \ref{Sec:KV}). We then explain the lightweight KV abstract (Section \ref{Sec:abstract}) and discuss the dynamic three-tier pipeline finally (Section \ref{Sec:pipeline}). 
\subsection{Overview}\label{Sec:overview}
We present LeoAM, a framework designed for accelerating long-context LLM inference on a commodity GPU with lightweight KV management overhead and low data transmission latency. LeoAM introduces two key design principles: (1) Adaptively resizing KV cache chunks based on context sparsity to minimize the overhead of the frequent importance evaluation under the long-context scenario. (2) Reducing the data transmission required for identifying important tokens and employing a three-tier pipelined offloading mechanism to reduce transmission latency. 
\begin{figure}[tbp]
    \centering
    \includegraphics[width=0.5\textwidth]{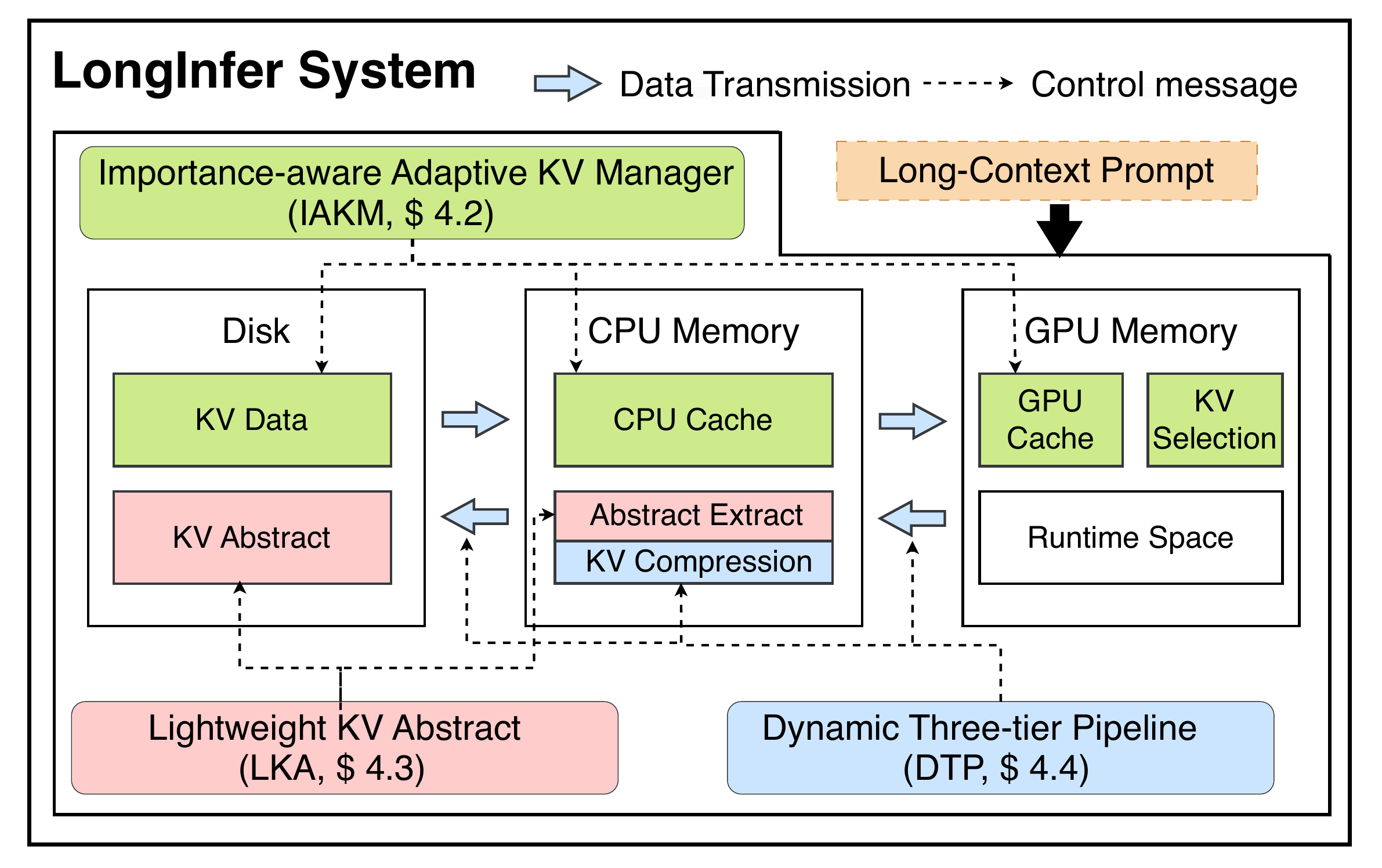}
    \caption{Overview of LeoAM System.}
    \label{fig:overview}
\end{figure}

\noindent\textbf{Overview.} Figure \ref{fig:overview} shows the overview of LeoAM. KV data is stored and distributed among the GPU cache, CPU cache, and the disk to address the severe memory constraints inherent to commodity hardware. Upon importance evaluation, KV data is selected and partitioned in a mutually exclusive manner across the GPU and CPU caches based on their assessed importance, with non-critical entries evicted to the disk for deferred access. To address the issue of the disk's limited I/O bandwidth, KV data is abstracted during storage to reduce the data volume required for the subsequent importance evaluation. Furthermore, KV data is compressed during transmission to further reduce transmission latency. The runtime memory is reserved for model parameters and intermediate inference artifacts, etc. LeoAM consists of three control modules: Importance-aware Adaptive KV Manager (IAKM, \$ \ref{Sec:KV}), Lightweight KV Abstract (LKA, \$ \ref{Sec:abstract}), and Dynamic Three-tier Pipeline (DTP, \$ \ref{Sec:pipeline}). IAKM is responsible for identifying important tokens with low overhead for efficient KV selection. LKA is designed to reduce transmission latency during importance evaluation under the GPU-CPU-Disk offloading framework. DTP fully exploits parallelism and dynamically controls the KV compression to minimize overall inference latency. 

LeoAM adopts the sum of each column in the attention weight matrix as the token importance metric, following a strategy similar to \cite{zhang2023h2o}. LeoAM also supports other token importance metrics proposed in recent works \cite{lee2024infinigen,zhao2024alisa,sheng2023flexgen,liu2024scissorhands}. Under chunked KV management, we use the upper bound of token importance within a chunk to represent the chunk’s overall importance, similar to the approach in \cite{tangquest}.

\noindent\textbf{Workflow.} In detail, during the prefilling stage, the system completes the computation and stores the locally required portion of the generated KV data in the GPU and CPU, while the remaining KV data is partitioned into initial chunk sizes and stored on the disk. Meanwhile, LKA abstracts the KV data on disk and stores the abstract back onto the disk as well. During the decoding stage, since token importance is query-dependent, each decoding step requires re-evaluation of token importance. The IAKM module adaptively partitions the KV data stored in the GPU and CPU with minimal overhead while performing importance evaluation. LKA extracts the KV abstract, and then the CPU evaluates the importance of the KV data stored on the disk based on it. Afterwards, the evaluation results are aggregated and used for KV selection, enabling the subsequent decoding computation. Throughout the LLM inference process, DTP performs streaming compression and transmission of data to reduce communication latency, while pipelining transmission and computation according to the characteristics of the three-tier cache hierarchy.

\subsection{Importance-aware Adaptive KV Management}\label{Sec:KV}
In importance-aware KV offloading for LLM inference, existing approaches typically manage KV data in a static manner, which leads to unnecessary computation and transmission overhead (Challenge 1). Building on our insights, we propose an importance-aware adaptive KV (IAKM) management strategy. The core idea is to adaptively determine the chunk size based on the density of the "attention desert" and manage the chunks in a tree structure. Chunks are merged or split according to token importance, enabling adaptive granularity for efficient KV management and importance evaluation. Since this strategy reduces the number of importance evaluation and resets the precise chunk size for important KV data, it significantly reduces both evaluation and transmission overhead. 

\begin{figure}[tbp]
    \centering
    \includegraphics[width=0.49\textwidth]{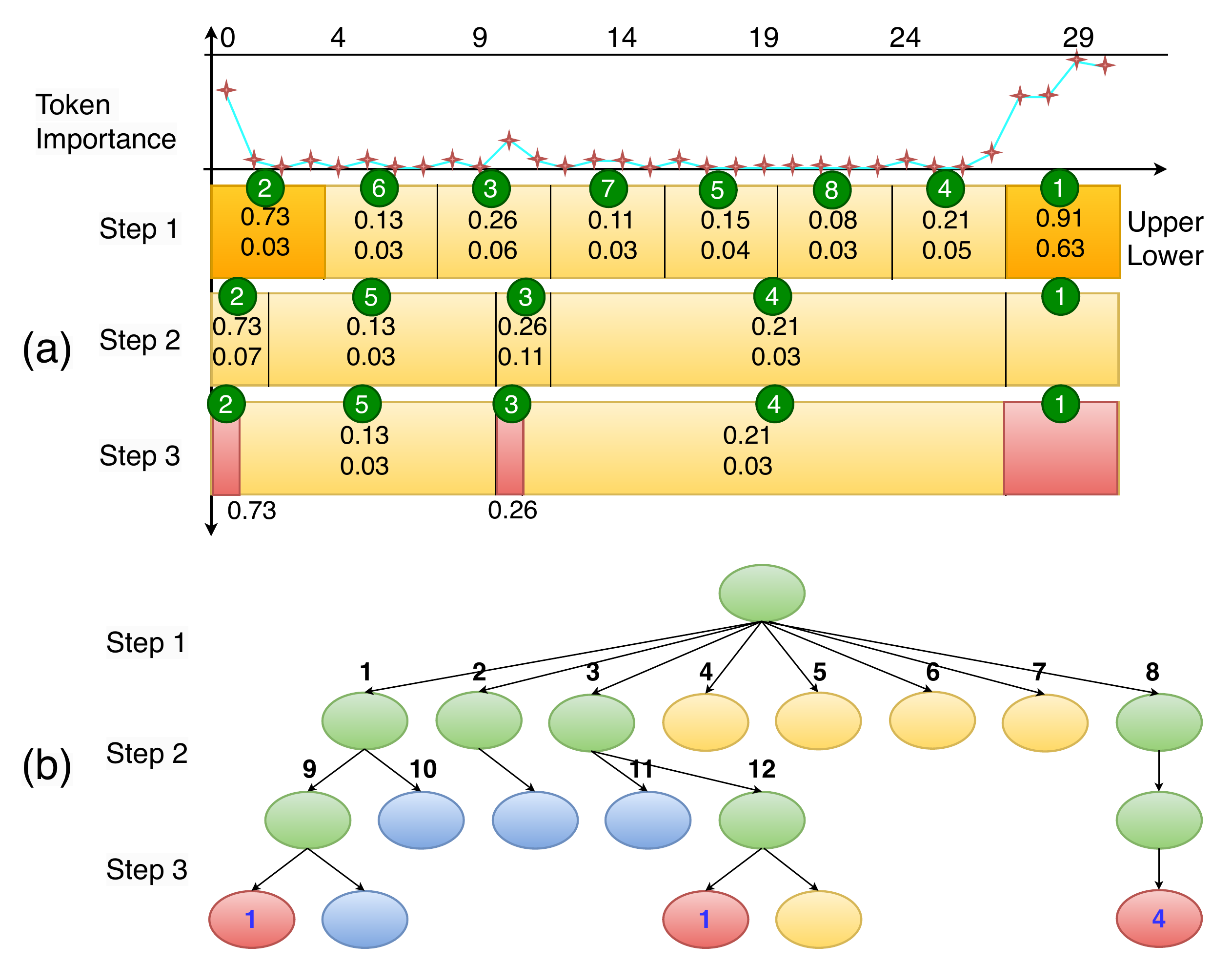}
    \caption{Tree-structured KV chunk management: (a) The orange and red chunks mean the identified important chunks without and with this method. (b) Chunk merge and split operations are the tree form of the tree construction process.}
    \label{fig:Tree-kv}
\end{figure}
\noindent\textbf{Tree-structured KV chunk management.} 
To better illustrate our design, Figure \ref{fig:Tree-kv} presents the adaptive KV chunk management and importance evaluation process within the IAKM module. Suppose there are $N = 32$ context tokens with $20\%$ (i.e., $6$) important tokens, and the token importance scores for the current decoding step are shown at the top of Figure \ref{fig:Tree-kv} (a). The tokens are divided into $m=8$ chunks, and chunk sizes are initiated as $s_i = 4, i=0,1,2,...,m-1$. When not using the tree-structured KV chunk management, existing approaches typically manage and evaluate KV data either at the token level or at a fixed chunk level: (1) In chunked management, as shown in Figure \ref{fig:Tree-kv}, Quest-like \cite{tangquest} methods require $8$ importance evaluations. However, the correct transmission ratio among the identified important KV data (i.e., the red region as a proportion of the orange region in the figure) is only $\frac{5}{8} = 62.5\%$, and it also misses the important KV data in the middle of step 3 (highlighted in red). (2) In token-level management, H2O-like \cite{zhang2023h2o} methods, require $32$ importance evaluations resulting in significant evaluation overhead. 
\begin{figure}[tbp]
    \centering
    \includegraphics[width=0.5\textwidth]{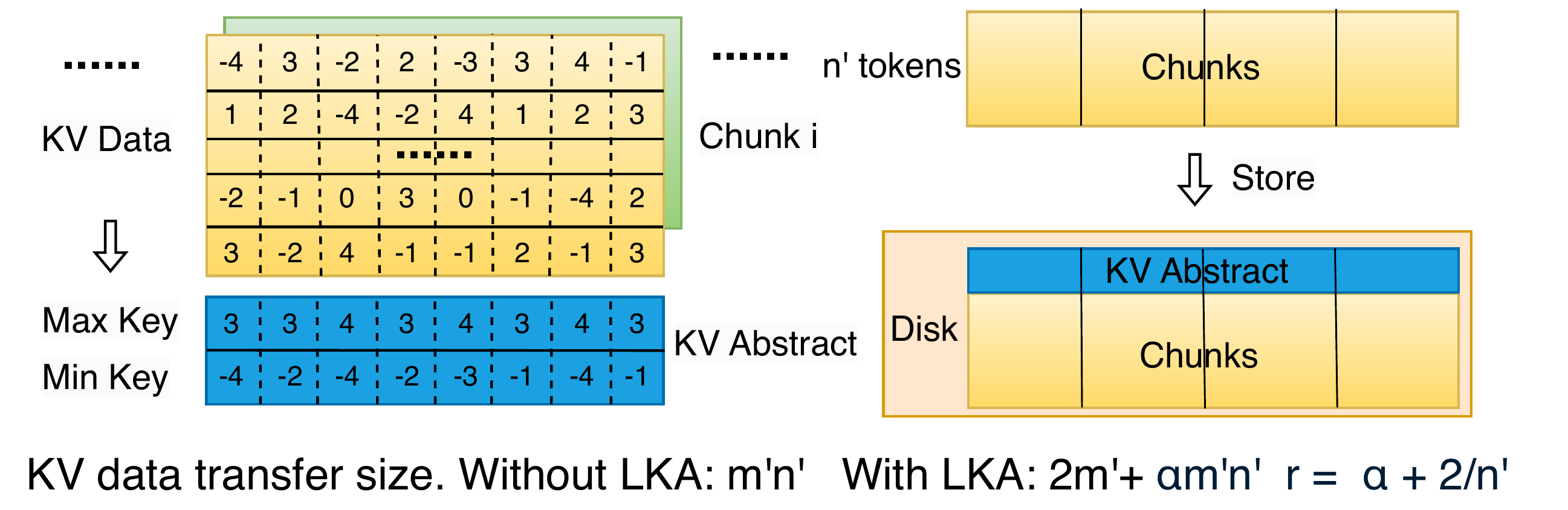}
    \caption{Lightweight KV Abstract: Suppose there are $m^{\prime}$ chunks with $n^{\prime}$ chunk size.}
    \label{fig:LKA}
\end{figure} 

In the tree-structured KV chunk management, we maintain a priority queue, where chunks are approximately ordered based on their importance. As shown in Figure \ref{fig:Tree-kv} (b), chunk merge and split operations form the tree construction process according to this queue. As shown in Figure \ref{fig:Tree-kv} (a), in Step 1, we first compute the upper and lower bounds for each initial chunk, then insert the corresponding values into a priority queue sorted by the upper bound, in which the green number indicates the sorted index of chunks. We then look up the queue from the front to progressively construct the confirmed set of $6$ important tokens. We define $Chunk^j_i$ as chunk $i$ in Step $j$ to distinguish chunks at different stages. According to the upper bound and lower bound of the chunks, there are at least $4$ important tokens in $Chunk^1_7$, $1$ token in $Chunk^1_0$, and $1$ token in $Chunk^1_2$. There are unimportant tokens in $Chunk^1_0, Chunk^1_2$ according to the lower bounds, and the continuous unimportant tokens can be merged. In Step 2, we split $Chunk^1_0$ and $Chunk^1_2$ and merge the unimportant parts and $Chunk^1_1$ into $Chunk^2_1$. $Chunk^1_3\sim Chunk^1_6$ are merged into $Chunk^2_3$. Then we continue to evaluate the upper and lower bounds of the split chunks. Following the same procedure in Step 2, we finally identify the important tokens in chunks. As shown in Figure \ref{fig:Tree-kv} (b), we eventually partition the tokens into $5$ chunks. Among them, the $3$ red chunks correspond to important tokens, with sizes of $1, 1, 4$, respectively. The blue and yellow chunks represent unimportant tokens, forming the so-called “attention desert”. 

Our method requires only $12$ importance evaluations, significantly reducing the evaluation overhead compared to $32$ evaluations in existing approaches. Moreover, in this case, we achieve a $100\%$ transmission ratio, improving upon the $62.5\%$ ratio observed in existing works.


\noindent\textbf{Dynamic chunk resizing.}
In our Insight 2, we observe that the token sparsity in the early two layers of the model and early few decoding steps is relatively low, with less continuity in the “attention desert” distribution. Applying our tree-structured chunk management technique at these layers may result in constructing an almost complete tree, which could unintentionally increase the number of importance evaluations. To tackle this issue, we introduce a dynamic chunk resizing strategy that adjusts the initial chunk size of KV data based on the token sparsity characteristics at each layer, thereby reducing unnecessary chunk evaluations. 

Specifically, we can leverage theoretical analysis to guide our experimental design. Let $\rho(l)$ denote the density of important tokens at layer $l$, and $A(m)$ denote the number of token evaluations. According to the design of tree-structured KV chunk management, at each tree level, there are $2$ times evaluations at probability $\rho(l)$, and each chunk with size $\frac{n}{m}$ can be split at most $log_2(\frac{n}{m})-1$ times. We can get:
\begin{align}
f(m) =& m + m2\rho(l) + m(2\rho(l))^2 + ...\\ \nonumber
=&m\sum_{i=0}^{log_2(\frac{n}{m})-1}(2\rho(l))^i,
\end{align}
In order to minimize $A(m)$, we compute its difference:
\begin{align}
\Delta A(m) = \sum_{i=0}^{log_2(\frac{n}{m})-1}(2\rho(l))^i - \sum_{i=log_2(\frac{n}{m})}^{log_2(\frac{n}{m-1})-1}(2\rho(l))^i
\end{align}
Following the standard approach of finding extrema via differences, we can approximately determine the value of $m$ that minimizes $A(m)$. Therefore, based on prior offline knowledge of the approximate density of important tokens at each layer, we can determine $m$ using the above formulas in the early two and early few decoding steps.
\begin{figure}[tbp]
    \centering
    \includegraphics[width=0.5\textwidth]{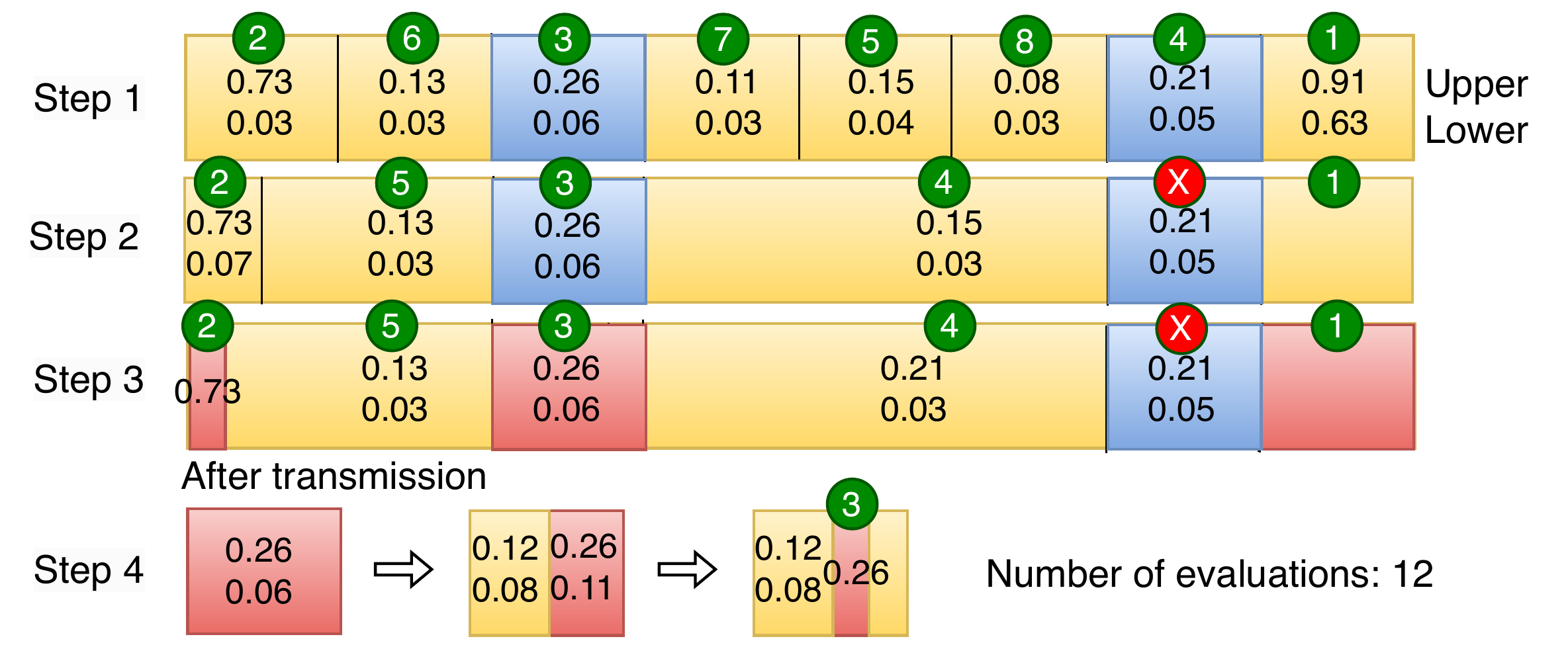}
    \caption{KV Management under LKA: The yellow and blue chunks represent the KV data stored in the CPU/GPU and KV abstract on the Disk, respectively.}
    \label{fig:LKA-2}
\end{figure}
\begin{figure}[tbp]
    \centering
    \includegraphics[width=0.5\textwidth]{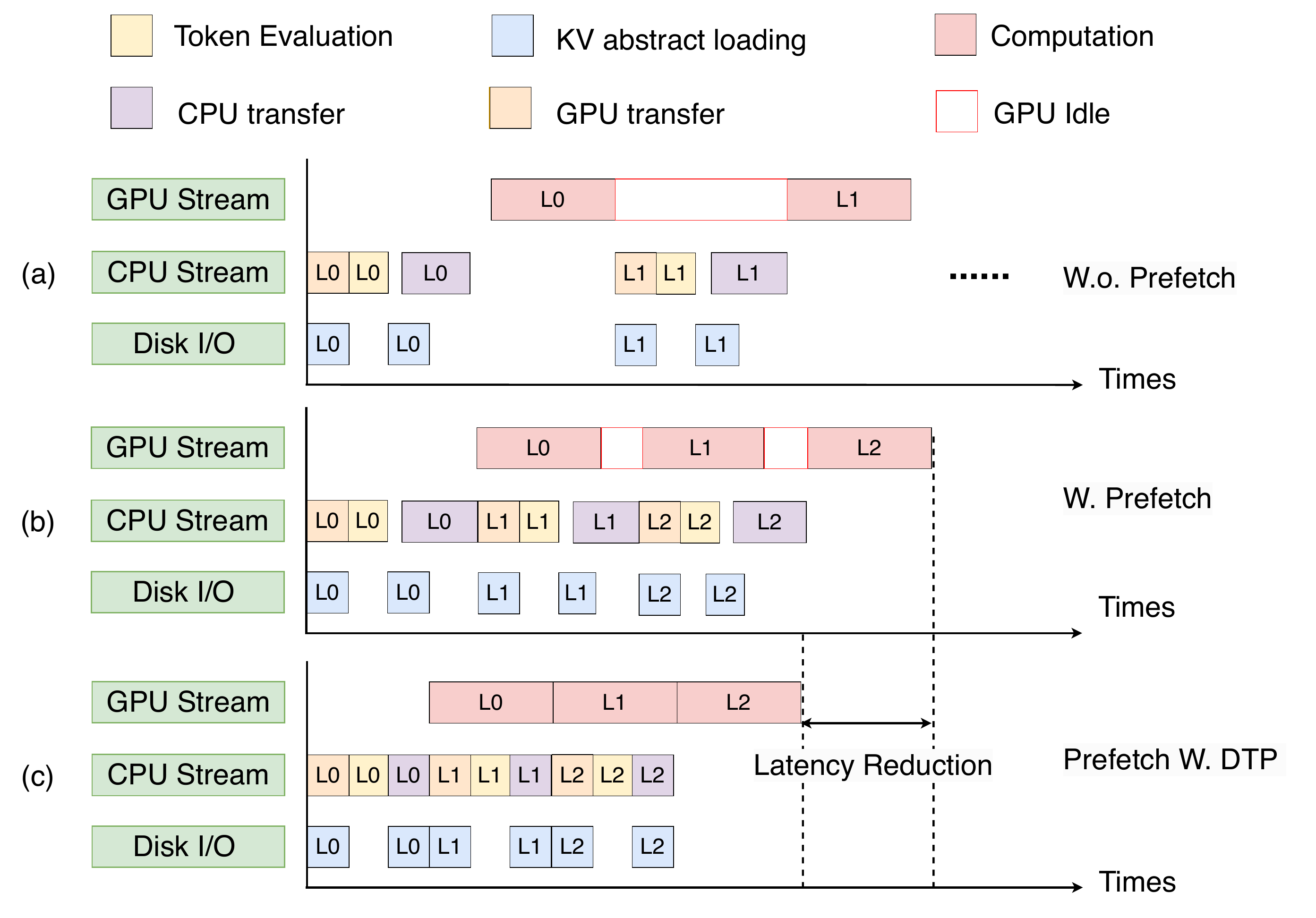}
    \caption{The latency in the decoding stage without prefetching and with prefetching and DTP techniques for 3 layers. $L_x$ represents $x^{th}$ layer.}
    \label{fig:DTP}
\end{figure}
\subsection{Lightweight KV Abstract}\label{Sec:abstract}
Due to the limited bandwidth of disks, KV data transmission incurs significant latency (challenge 2). To reduce the volume of KV data transmitted from the disk, we propose the Lightweight KV Abstract (LKA) mechanism. The core idea of LKA is to extract a compact representation of KV data features required for importance evaluation as KV abstract, based on the computational characteristics of the evaluation algorithm. During token importance evaluation, only the KV abstract needs to be transmitted to represent the corresponding KV data chunk on disk.

\noindent\textbf{KV abstract extraction.} Figure \ref{fig:LKA} shows the process of LKA. We construct the KV abstract of a chunk by extracting its element-wise maximum and minimum key vectors. Given that the disk provides ample storage but suffers from low I/O bandwidth, we can store the KV abstract alongside the full KV data on disk. When evaluating the KV data residing on disk, we transmit its lightweight abstract rather than the entire data. Using these abstracted keys along with the query at the current decoding step, we can efficiently compute the chunk’s upper and lower bounds. Without the LKA mechanism, all KV data corresponding to the $n^{\prime}$ tokens have to be transmitted during importance evaluation. In contrast, with LKA, each chunk only transmits the two tokens' Key data. Assuming there are $m^{\prime}$ chunks and a proportion $\alpha$ of tokens are important on disk, the total data transmitted is approximately $2m^{\prime} + \alpha m^{\prime}n^{\prime}$, which accounts for only a fraction of $r = \alpha + \frac{2}{n^{\prime}}$ of the original data volume. ($\alpha$ and $\frac{2}{n^{\prime}}$ are values marginally above zero, e.g., $\alpha = 0.1$, $n^{\prime} = 32$, $r = 13.25\%$)

\noindent \textbf{KV Management and optimization under LKA.} LKA can only transmit $r = \alpha + \frac{2}{n{\prime}}$ proportion of all kV data on disk. This implies that further optimization of KV data management can be achieved by reducing $\alpha$. Since we observe that the token sparsity in the early few layers is relatively low, we do not store their KV data on disk. Instead, we apply only a two-tier offloading between the GPU and CPU for these layers. In addition, we maintain an access frequency table for all tokens and avoid offloading KV data with high access rates to the disk. These two strategies together can reduce $\alpha$, thereby decreasing the proportion of KV data that needs to be accessed from the disk. Additionally, we cache all KV data replicas on disk to avoid additional I/O access latency when less important chunks are evicted from the CPU to disk.

Figure \ref{fig:LKA-2} shows the KV management under LKA. Unlike IAKM performed on GPU/CPU, the KV abstract cannot be split; the chunk $C^1_2$ represented by its KV abstract remains unchanged throughout the importance evaluation process. In contrast, chunk $C^1_6$ represented by its KV abstract, is identified as unimportant in Step 2 and thus excluded from subsequent chunk merging and transmission from the disk. Ultimately, $C^1_2$ is selected as the candidate important chunk and needs to be fetched from disk. Step 4 is then split to enable the selection of finer-grained important KV. 

\subsection{Dynamic Three-tier Pipeline}\label{Sec:pipeline}
In our importance-aware GPU-CPU-Disk offloading framework, we reduce end-to-end inference latency by combining the IAKM mechanism with system-level pipelining of computation and data transmission. To tackle the inherent latency of multi-tier KV offloading, we propose a Dynamic Three-tier Pipeline, whose core idea is to reduce the compute bubble of the GPU, i.e., increase the overlapping between computation and transmission. Additionally, we integrate a dynamic KV compression technique to further reduce the inference latency by minimizing the transmission overhead across tiers.

Under the GPU-CPU-Disk KV offloading, KV data is distributed across the GPU, CPU, and disk. KV data from all three tiers needs to be assessed and consolidated to evaluate token importance. Since the CPU serves as the bridge between the GPU and the disk, we centralize the evaluation process on the CPU and then transmit the important KV data to the GPU. 

As shown in Figure \ref{fig:DTP} (a), significant GPU idle time exists due to KV importance evaluation and data transmission. The existing KV prefetching technique between the GPU and CPU \cite{lee2024infinigen} can enable layer-wise overlap of computation and KV transmission, which provides a clear technical pathway: we can perform LLM inference computation for the current layer while simultaneously conducting KV importance evaluation and KV prefetching for the next layer. A key difference is that we also need to consider the prefetching of KV data from the disk. Figure \ref{fig:DTP} (b) shows the execution timeline with the prefetching technique. At each decoding step, we first transmit the offloaded KV data of layer 0 from the previous decoding step from the GPU to the CPU, while simultaneously loading the KV abstracts from the disk into the CPU. We then perform importance evaluation on the KV data and transmit the important chunks from both the disk (through CPU for KV Management and optimization under LKA) and the CPU back to the GPU for inference computation. During this computation, we concurrently evaluate and transfer the KV data for layer 1. This process continues iteratively until the computations of all layers are completed.

\noindent \textbf{Dynamic KV compression.}
In the ptrfetching technique, the problem is that the overall latency of the next layer’s KV data operations often cannot be fully hidden due to the longer KV transmission latency, resulting in unavoidable CPU idle periods, as shown in Figure \ref{fig:DTP} (b). We can leverage KV compression techniques to reduce this transmission latency. However, if all KV data is compressed, it can significantly reduce transmission latency but also incur additional decompression overhead on the GPU. Therefore, we need to dynamically determine an optimal compression ratio of KV data to minimize the overall latency, and trade off between transmission latency reduction and decompression computation.

We assume that the total KV data to be transmitted in this layer is $D$. A proportion $\theta$ of the data is compressed, with a compression ratio $\delta$. Let the PCIe bandwidth be $B$, and the decompression time for $D\theta$ compressed data be $t(D\theta)$, where $t(), B$ can be determined by an offline hardware test. The overall latency can only be fully hidden within the computation time when condition $T_0 + \frac{D(1-\theta)+D\theta\delta}{B} \leq T_c + t(D\delta)$ holds, where $T_0$ represents the other evaluation and transmission latency. Therefore, we dynamically determine the value of $\theta$ such that the equality condition holds, which minimizes GPU idle time, as illustrated in Figure \ref{fig:DTP} (c).
\begin{figure*}[tbp]
    \centering
    \includegraphics[width=\linewidth]{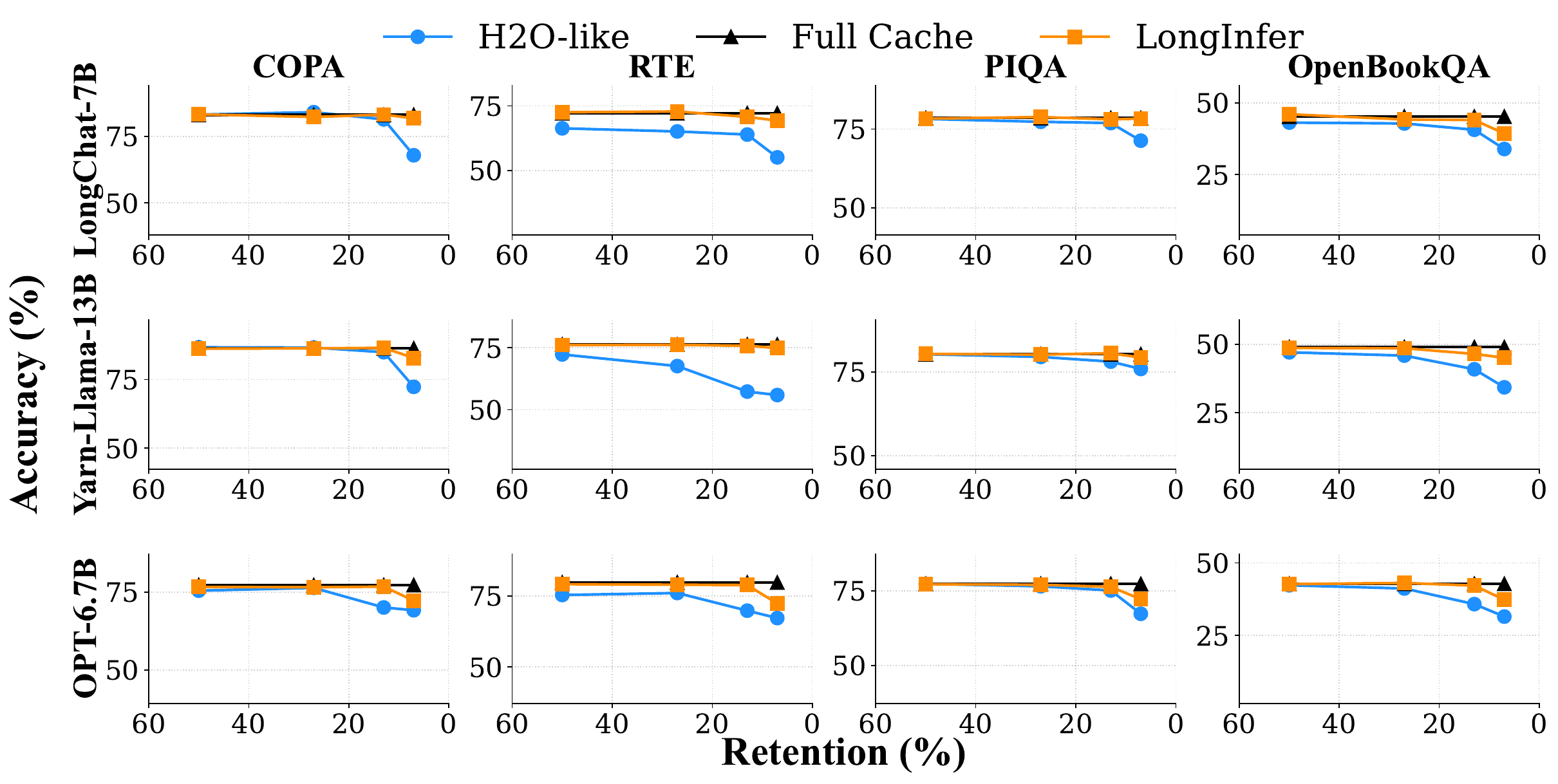}
    \caption{The accuracy of different systems across four datasets and three models under different relative KV Cache sizes.}
    \label{fig:3x4_accuracy}
\end{figure*}

\section{Implementation}
LeoAM builds upon the current offloading-based framework, FlexGen \cite{sheng2023flexgen}, by integrating over 1000+ lines of additional code to implement the proposed design. This choice was made because of the complete implementation of the model and KV data offloading in FlexGen, and we can modify and extend it to realize our proposed designs. For essential long-text data processing and experiments related to the insights, we conducted inference analyses in the preliminary experiments using the Hugging Face Transformers framework \cite{wolf2019huggingface}. LeoAM is capable of accelerating long-context inference for most open-source models, such as Llama \cite{touvron2023llama}, LongChat \cite{longchat}, and the models from LongLora \cite{chenlonglora}, etc., provided that the models themselves support sufficiently long-context lengths. 

\section{Evaluation}\label{Sec:6}
\subsection{Experimental Setup}
\textbf{Hardware.} To better align with the practical needs of most users who rely on their personal PCs, we select a GPU setup, consisting of an RTX 4090 GPU with an Intel Core(TM) i7-14700K CPU, 24 GB of GPU memory, 120 GB of host memory, 800 GB Intel SSD, and a PCIe 4.0 interface. The measured read throughput of SSDs is around 7GB/s. To emulate commodity GPUs with varying memory capacities, we constrain the available GPU and CPU memory by controlling the proportions of KV data size in CPU and Disk.

\noindent \textbf{Models.} In existing open-source models like LLaMA \cite{touvron2023llama}, support for extended context lengths is limited. To address this, we curated a selection of models that have been fine-tuned on the original architectures to accommodate long-context scenarios. We use the longchat-7B-v1.5-32k \cite{longchat7b}, and yarn-llama-2-13B-128k \cite{yarnllama2_13b}. We also use the OPT models to test the performance (e.g., OPT-6.7B). Throughout the inference process, all KV Cache data was stored in FP16 quantization mode and compressed in INT4.

\noindent \textbf{Datasets.} First, to evaluate the accuracy preservation of LeoAM, we selected four few-shot tasks from the lm-evaluation-harness benchmark \cite{chen2019few} (OpenBookQA, PIQA, RTE, and COPA) and PG-19 \cite{rae2019compressive}. Since our focus is on long-context inference, we also sample a variety of prompts from LongBench \cite{bai2023longbench} to evaluate the acceleration performance of LeoAM. 

\noindent \textbf{Baselines inference framework.} 
We compare LeoAM with four baselines, including: (1) H2O-like: We adopt a method similar to H2O \cite{zhang2023h2o} for token importance evaluation in the token level, which loads only the KV data of important tokens based on the evaluation results.	(2)	H2O-like-chunked: Based on the H2O-like method, this method evaluates token importance at the chunk level to reduce evaluation overhead. (3) Prefetch-based: Similar to InfiniGen \cite{lee2024infinigen}, this method adds the prefetching technique on top of importance-aware selection to increase the overlap between computation and transmission. In the baselines, we need to transmit all of the KV data from disk for importance evaluation.


\noindent \textbf{Comparison metics.}
First, when evaluating model accuracy, we adopt the accuracy (\%) metric as used in prior work \cite{zhang2023h2o}. Second, for inference latency and acceleration performance, we follow previous studies \cite{lee2024infinigen} by measuring inference latency (s) and throughput (tokens/s). Additionally, during importance evaluation and KV selection, we default to loading only 10\% of the most important KV data. For the first two transformer layers, however, we load 50\% of the KV cache by default. We uniformly set the initial chunk size as $64$ tokens in baseline and LeoAM and resize the chunk size as $8$ in the early two layers and decoding steps. To prevent out-of-memory errors, especially when model weights are large, we consistently store model weights across both the CPU and GPU.

\subsection{Inference Performance Analysis}\label{Sec:7.2}
\noindent \textbf{Model generation accuracy.} We test the end-to-end inference performance of LeoAM compared against H2O-like and Full cache on our device. The accuracy of the prefetch-based and H2O-chunked methods is comparable to that of the H2O-like method. Therefore, we only compare against the H2O-like and Full cache baselines for simplicity. Figure \ref{fig:3x4_accuracy} shows the comparison of model generation quality (accuracy \%) between LeoAM and these baseline inference systems under different relative KV Cache sizes (\%). Across different models and datasets, we observe that LeoAM incurs less than a 1\% accuracy drop compared to the baselines. Moreover, we find that smaller relative KV Cache sizes lead to larger accuracy degradation, highlighting the need to determine an appropriate proportion for the relative KV Cache size. 

\begin{figure*}[tbp]
    \centering
    \includegraphics[width=\linewidth]{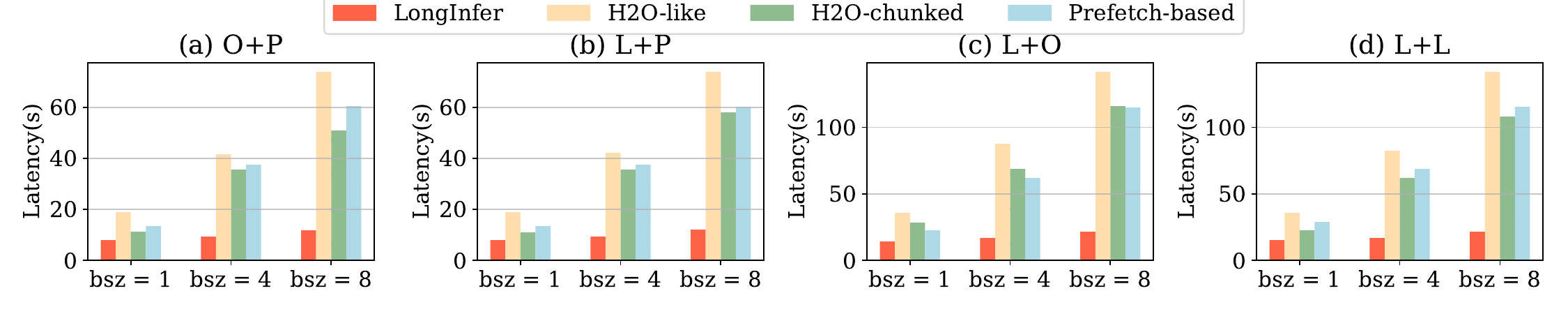}
    \caption{The inference latency cooperation with baselines of different systems across two datasets and three batch sizes. O: OPT-6.7B, L:LongChat-7B, P:PG-19, L:LongBench.}
    \label{fig:1x4_latency}
\end{figure*}
\begin{figure}[tbp]
    \begin{minipage}[t]{0.50\linewidth}
        \centering
        \includegraphics[width=\textwidth]{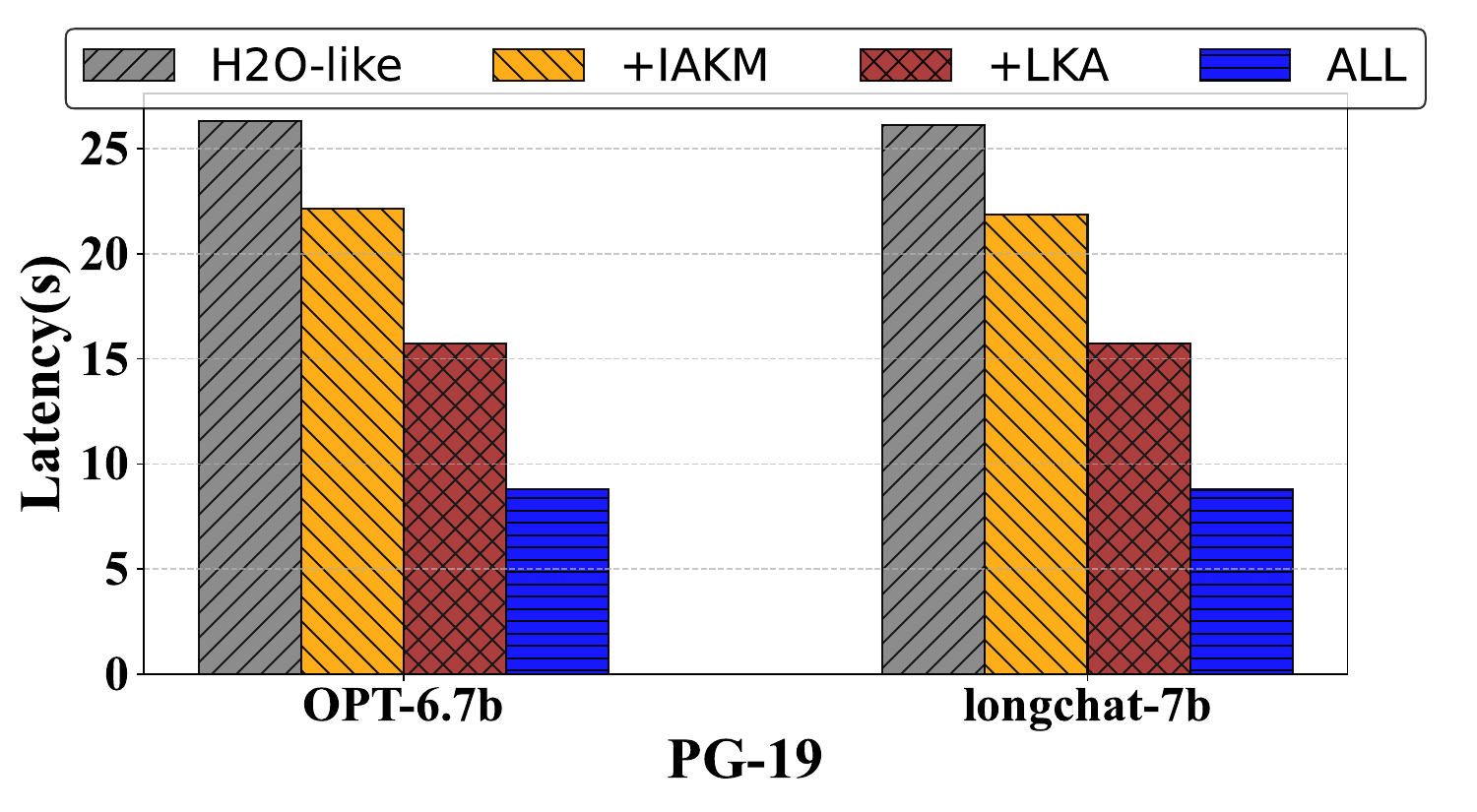}
        \centerline{(a) PG-19.}
    \end{minipage}%
    \begin{minipage}[t]{0.50\linewidth}
        \centering
        \includegraphics[width=\textwidth]{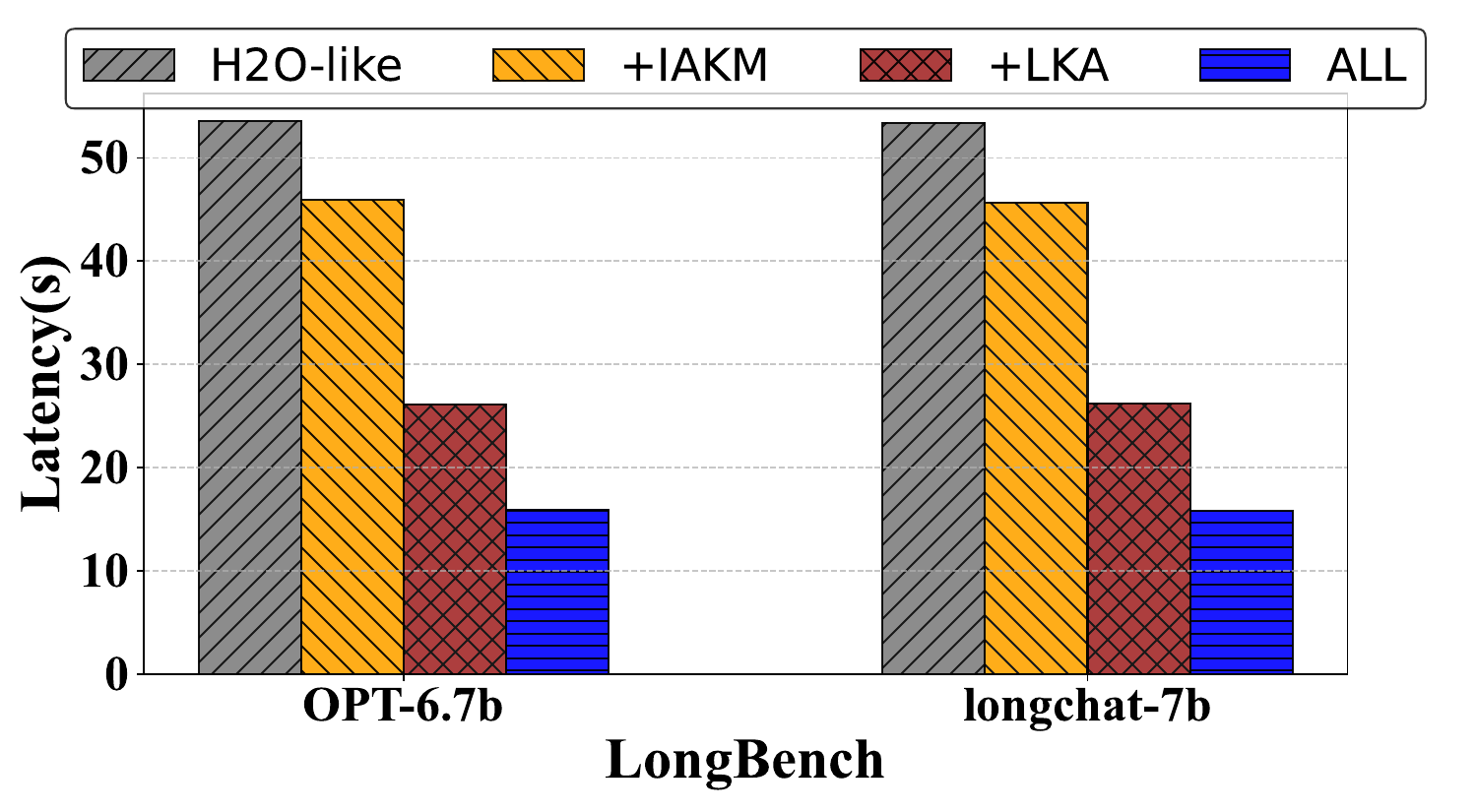}
        \centerline{(b) LongBench.}
    \end{minipage}
    \caption{The individual technique breakdown of latency performance.}
    \label{fig:inditival_latency}
\end{figure}

\noindent \textbf{Inferecne Latency.}
We conduct inference using LongChat-7B and OPT-6.7B models on LongBench and PG-19 datasets with 128 output tokens and batch size = 1,4,8. Figure \ref{fig:1x4_latency} illustrates the inference latency (including prefilling and decoding stages) between LeoAM and the baselines. It shows LeoAM reduces inference latency by $3.46\times$ compared to the better baseline on average. As the batch size increases, the size of the KV data increases and our acceleration becomes more pronounced. At a batch size of 8, LeoAM achieves a $5.47\times$ speedup. 


\subsection{Individual Technique Breakdown}
We evaluate the latency performance contributions of individual techniques using OPT-6.7B and LongChat-7B on the PG-19 and LongBench datasets. We use the H2O-like method as the baseline, which performs token-level importance evaluation. +LKA introduces the LKA module, where only KV abstracts are transferred from disk during token importance evaluation. +IAKM further incorporates the IAKM module, which significantly reduces the overhead of the evaluation process through a lightweight mechanism. ALL represents the full LeoAM system with all components enabled. In these experiments, we set the importance rate to 0.1 and use a batch size of 2. Figure \ref{fig:inditival_latency} shows the latency improvements brought by each optimization across different models and datasets. We observe that each optimization contributes to a reduced latency, with the ALL configuration achieving the lowest latency. This demonstrates the effectiveness of LeoAM in reducing inference delay. Moreover, the performance gains vary across different datasets and models, highlighting the adaptability of each individual technique of LeoAM to diverse inference scenarios.

\begin{figure}[tbp]
    \centering
    \includegraphics[width=\linewidth]{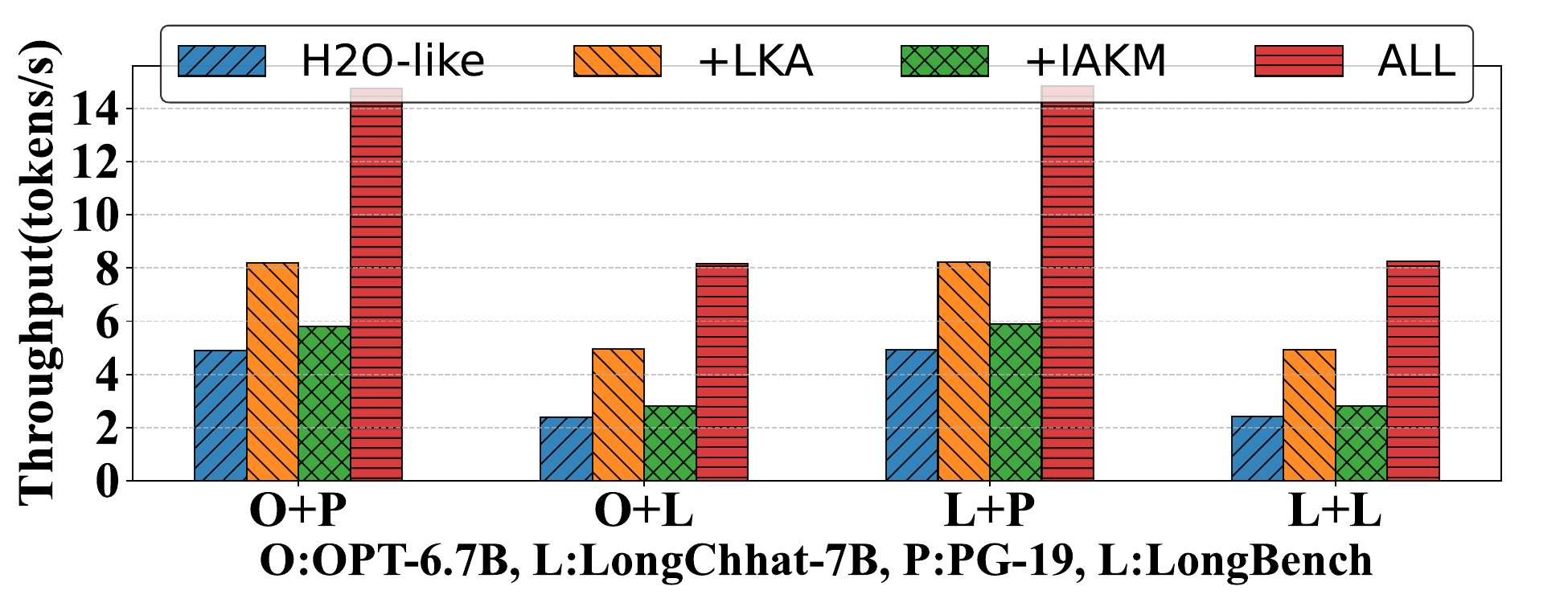}
    \caption{The throughput between different models and datasets.}
    \label{fig:individual_throughput}
\end{figure}

Figure \ref{fig:individual_throughput} shows the throughput improvements brought by each individual optimization under the same experimental settings. We observe that every single technique contributes to higher inference throughput, while the combination of all optimizations (ALL) achieves the highest throughput performance, demonstrating LeoAM’s efficient inference capabilities and acceleration effectiveness. Furthermore, the effectiveness of each individual optimization differs across datasets, underscoring the effectiveness and adaptability of each individual LeoAM technique in varying inference scenarios. For example, as shown in Figure \ref{fig:inditival_latency}, on the PG-19 and LongBench datasets, the IAKM, LKA, and ALL configurations improve latency performance by 15.8\%, 40\%, and 66\%, and by 14\%, 51.2\%, and 70\%, respectively. In Figure \ref{fig:individual_throughput}, on the PG-19 and LongBench datasets, the IAKM, LKA, and ALL configurations achieve throughput improvements of $1.19\times$, $1.67\times$, and $3.02\times$, and $1.17\times$, $2.06\times$, and $3.40\times$, respectively.

\subsection{Sensitivity Analysis}

\begin{figure}[tbp]
    \centering
    \includegraphics[width=\linewidth]{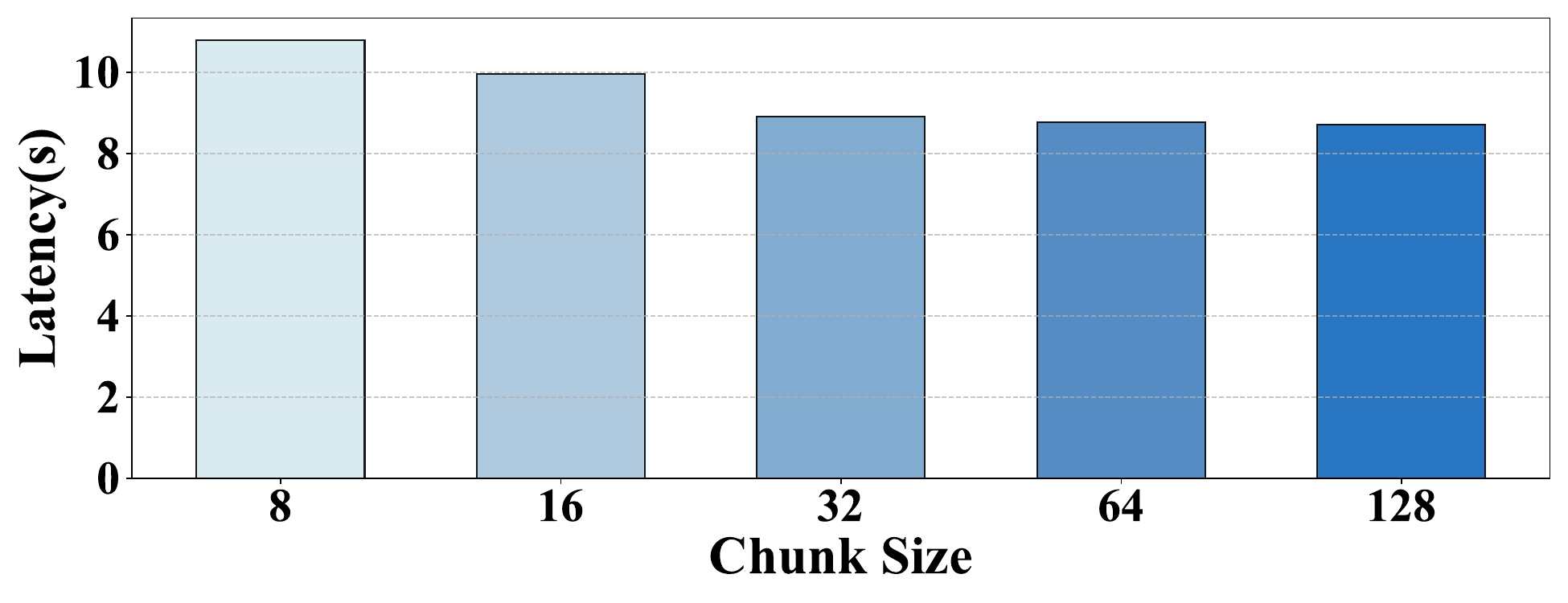}
    \caption{The latency performance between different chunk sizes.}
    \label{fig:latency_chunk_size}
\end{figure}

\noindent\textbf{Chunk Size.} Figure \ref{fig:latency_chunk_size} illustrates the latency performance across different chunk sizes. We conduct the experiments on the OPT-6.7B model with an output length of 128 tokens and an importance rate of 0.2. We observe that as the chunk size increases, the latency consistently decreases and the rate of reduction gradually diminishes. Notably, when the chunk size increases from 64 to 128, the latency decreases by approximately 0.8\%. This is because when the chunk size is small, the overhead from importance evaluation and KV abstract transmission becomes significant, leading to higher latency. As the chunk size increases, both types of overhead are reduced. However, once the chunk size exceeds a certain threshold (e.g., 64), the cost of importance evaluation begins to rise, even though the transmission overhead of KV abstracts continues to decrease. Therefore, considering this trade-off, we select 64 as the default chunk size in our experiments. 

\begin{figure}[tbp]
    \centering
    \includegraphics[width=\linewidth]{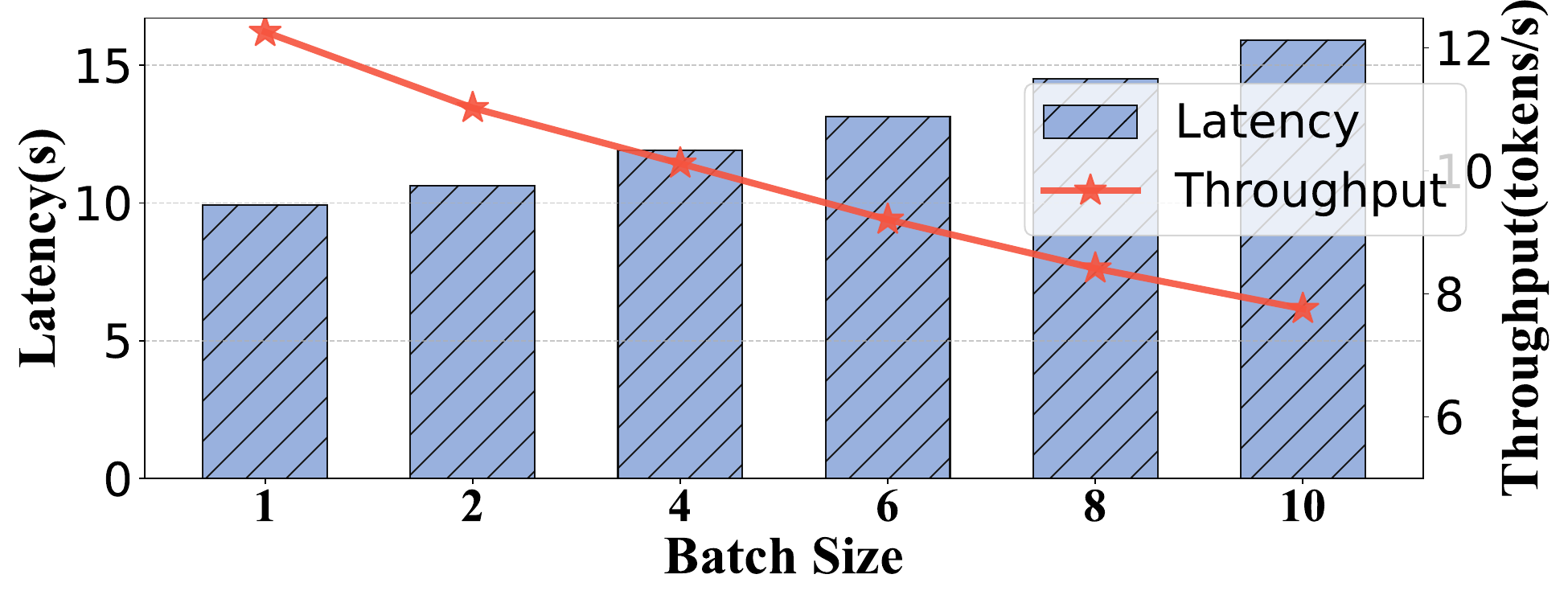}
    \caption{The latency and throughput performance between different batch sizes.}
    \label{fig:latency_batch_size}
\end{figure}

\noindent\textbf{Batch Size.} Figure \ref{fig:latency_batch_size} presents the latency and throughput performances under different batch sizes. We set the output length to 128 tokens, the importance rate to 0.1, and the chunk size to 64. We observe that as the batch size increases, the latency increases while the throughput decreases. This is because the size of the KV Cache grows proportionally with the batch size, resulting in a larger volume of KV data that needs to be transferred and computed, thereby introducing additional transmission and computation delays both in prefilling and decoding stages.

\subsection{Overhead analysis}
\noindent\textbf{Time overhead.} In LeoAM, we need to compute and load KV abstracts from disk for token importance estimation. Although it brings additional computation and transmission overhead, it accounts for only 1.56\% of the overall system overhead on average, due to the small size of KV Abstracts when the chunk size is set to 64. Furthermore, our IAKM module adopts a more lightweight evaluation strategy, reducing overhead by $15\%\sim41\%$ compared to traditional token-level importance assessment methods.

\noindent\textbf{Space overhead.} KV abstracts are stored alongside KV data on disk, which slightly increases the storage footprint. When the chunk size is 64, each chunk only brings additional data equivalent to that of a single token's KV data, resulting in less than $1.6\%$ memory increase in overall KV storage. Moreover, during token importance evaluation, a tree structure is maintained, including KV abstracts and upper/lower bounds, etc., which also incur minimal overhead. For a chunk size of 64 and an importance rate of $0.2$, this structure requires, on average, only $2.2\%$ of the total KV data size, which is a negligible cost given the relatively large capacity of disk and CPU memory.
\vspace{-0.1in}

\section{Related Work}
\textbf{KV Caching operations for LLM inference.}
KV Caching technique was first introduced by Reiner Pope et al. \cite{pope2023efficiently}, which reduces computational complexity and has recently been widely adopted in the field of long-context LLM inference acceleration \cite{luohekeep}, etc. KV Caching operations during the pre-filling stage are crucial for long-context LLM inference, including compression \cite{jhacharacterizing,dongget,liu2024scissorhands,ge2023model,wan2024look,yang2024pyramidinfer,gemodel}, selection \cite{huang2024locret,tangquest,adnan2024keyformer,jiangminference,liu2024retrievalattention,zhu2024near,kwon2023efficient}, and pruning \cite{fulazyllm,xu2024think,chokv}, and quantization \cite{hooper2024kvquant,dong2024qaq,tan2024alignedkv} techniques. LeoAM focuses on leveraging the KV Cache offloading technique to enable long-context inference on a commodity.\\
\textbf{Long-Context LLM Inference System.} Most of the existing works primarily focus on implementing long-context LLM inference services in the cloud 
\cite{lin2024infinite,hooper2024kvquant,agrawal2024mnemosyne,wu2024loongserve,lim2024accelerating} or on industrial-grade GPUs \cite{sheng2023flexgen,zhao2024alisa,pan2024instinfer,zhang2024pqcache,liu2024cachegen,yao2024cacheblend}. In data centers, \cite{lin2024infinite} and \cite{hooper2024kvquant} mainly focus on distributed KV Caching and attention layers scheduling and quantization. \cite{agrawal2024mnemosyne} and \cite{wu2024loongserve} primarily investigated how to accelerate LLM inference through sequential parallelism, aiming to enhance LLM service capabilities. In industrial GPUs, \cite{zhao2024alisa},\cite{zhang2024pqcache}, and \cite{liu2024cachegen} explore the sparsity, quantization, and compression of KV Cache, respectively, to accelerate long-context LLM serving. \cite{pan2024instinfer} mainly proposes an LLM
inference system considering both attention offloading and KV Cache transmitting optimization. However, the works above primarily serve cloud-based or industrial-grade GPUs, making it challenging to address the limitations of a resource-constrained commodity GPU. Some researchers have explored deploying LLMs on a commodity GPU or edge device \cite{song2023powerinfer,huang2024edgellm}; however, their focus has not been on long-context inference scenarios. We fill this gap. \\
\textbf{Importance-aware KV management for LLM Inference} 
Recent studies \cite{zhang2023h2o,zhao2024alisa,liu2024scissorhands,lee2024infinigen,tangquest,gao2024cost,juravsky2024hydragen,ye2024chunkattention,zheng2024sglang} have focused on leveraging token sparsity to optimize long-context LLM inference. These approaches typically evaluate the importance of each token’s KV data, retain the important KVs in GPU memory, and evict or offload the unimportant KV data to CPU memory. Most of these methods are built upon a two-tier GPU-CPU offloading framework for KV management. However, under the resource constraints of a commodity GPU, LeoAM introduces a three-tier KV offloading framework spanning GPU, CPU, and disk to enable importance-aware KV management. Moreover, existing works typically rely on static KV management at the token \cite{zhang2023h2o,zhao2024alisa,liu2024scissorhands,lee2024infinigen} or chunk \cite{tangquest,gao2024cost,juravsky2024hydragen,ye2024chunkattention,zheng2024sglang} level, which leads to substantial computation and communication overhead. LeoAM proposes an importance-aware adaptive KV management mechanism to address the challenges above.

\section{Conclusion}
In this paper, we present LeoAM, a token-importance-aware inference system tailored for long-context LLMs on a single commodity GPU. By leveraging adaptive KV chunk management and lightweight KV abstract evaluation, LeoAM minimizes both computation and I/O overheads. It further reduces disk-induced latency through dynamic compression and a three-tier pipelined transmission-computation strategy. Built on an offloading-based framework, LeoAM demonstrates significant performance gains, achieving up to $3.46\times$ speedup on average in inference latency, while maintaining high output quality. Our work highlights a practical and efficient solution for enabling long-context LLM inference on a resource-constrain commodity CPU.
\bibliographystyle{ACM-Reference-Format}
\bibliography{sample}

\appendix









\end{document}